\documentclass[lettersize,journal]{IEEEtran}
\usepackage{amsmath,amsfonts}
\usepackage{algorithmic}
\usepackage{algorithm}
\usepackage{array}
\usepackage[caption=false,font=normalsize,labelfont=sf,textfont=sf]{subfig}
\usepackage{textcomp}
\usepackage{stfloats}
\usepackage{url}
\usepackage{verbatim}
\usepackage{graphicx}
\usepackage{cite}
\usepackage[dvipsnames]{xcolor} % code color
\usepackage{booktabs}                  % only used for the table example
\usepackage{array}
\usepackage{makecell}
\usepackage{lipsum}                    % used to generate placeholder text
\usepackage{mwe}                       % used to generate placeholder figures
\usepackage[colorinlistoftodos,textwidth=1.0cm,textsize=tiny]{todonotes}
\usepackage{authblk}
\usepackage{amsmath}
\usepackage{multirow}
\usepackage{xurl}
\usepackage{xspace}
\usepackage{algorithm}
\usepackage{algorithmic}
\usepackage{graphicx}
\usepackage[pagebackref,bookmarks]{hyperref}
\usepackage{cleveref} 
\usepackage{xr}
\usepackage{graphicx}
\usepackage{capt-of}
\usepackage{multicol}
\usepackage{mathptmx}                  % use matching math font
\usepackage{tabularx}

\usepackage{tcolorbox}
\tcbuselibrary{breakable,skins}
\usepackage{listings}
\usepackage{xcolor}
\input{glyphtounicode}
\newcommand{\name}{{WidgetVA}\xspace}
\newcommand{\bench}{{WidgetVABench}\xspace}
\newcommand{\eg}{\textit{e.g.}\xspace}
\newcommand{\ie}{\textit{i.e.}\xspace}

\newcommand{\etal}{\textit{et al.}\xspace}

\newcommand{\revise}[1]{\textcolor{black}{#1}}

\usepackage[T1]{fontenc}
\usepackage{mathptmx}
\usepackage{colortbl}
\usepackage{pgf}

\definecolor{singleheatblue}{RGB}{76,146,195}

\newcommand{\singleheat}[1]{%
  \pgfmathtruncatemacro{\singleheatlevel}{round(6+0.58*(#1))}%
  \edef\singleheatbox{%
    \noexpand\cellcolor{singleheatblue!\singleheatlevel!white}%
  }%
  \singleheatbox #1
}

\newcommand{\singlestatelabel}[1]{%
  \multirow{3}{*}{%
    \rotatebox{90}{\bfseries $#1$}%
  }%
}

\crefname{paragraph}{section}{sections}
\Crefname{paragraph}{Section}{Sections}
\begin{document}

\title{\revise{\name: A Widget-Centric Framework and Benchmark for Agentic Visual Analytics}}
\author{
Yutong~Chen\textsuperscript{\textdagger},
Zhike~Tang\textsuperscript{\textdagger},
Zhihao~Mai,
Zhihao~Shuai,
Danli~Luo,
Jing~Xu,
and~Weikai~Yang\textsuperscript{*}
\thanks{Y. Chen, Z. Tang, Z. Shuai, J. Xu, and W. Yang are with
the Data Science and Analytics Thrust, Information Hub,
The Hong Kong University of Science and Technology (Guangzhou),
Guangzhou, China.}

\thanks{
Z. Mai is with Nanyang Technological University, Singapore.}

\thanks{
D. Luo is with the School of Electronics and Information Technology
(School of Microelectronics), Sun Yat-sen University, Guangzhou, China.}

\thanks{
Y. Chen and Z. Tang contributed equally.
W. Yang is the corresponding author. Email: weikaiyang@hkust-gz.edu.cn}
}
% \author{
% Yutong~Chen\textsuperscript{1,\textdagger},
% Zhike~Tang\textsuperscript{1,\textdagger},
% Zhihao~Mai\textsuperscript{2},
% Zhihao~Shuai\textsuperscript{1},
% Danli~Luo\textsuperscript{3},
% Jing~Xu\textsuperscript{1},
% and~Weikai~Yang\textsuperscript{1,*}
% %
% \IEEEcompsocitemizethanks{
%     \IEEEcompsocthanksitem
%     \textsuperscript{1}Data Science and Analytics Thrust, Information Hub,
%     The Hong Kong University of Science and Technology (Guangzhou),
%     Guangzhou, China.

%     \IEEEcompsocthanksitem
%     \textsuperscript{2}School of Mechanical and Aerospace Engineering (MAE), Nanyang Technological University, Singapore.

%     \IEEEcompsocthanksitem
%     \textsuperscript{3}School of Electronics and Information Technology
%     (School of Microelectronics), Sun Yat-sen University, Guangzhou, China.

%     \IEEEcompsocthanksitem
%     \textsuperscript{\textdagger}Equal contribution.
%     \textsuperscript{*}Corresponding author.

\markboth{Transactions on Visualization and Computer Graphics}%
{Yutong Chen, Zhike Tang, \MakeLowercase{\textit{(et al.)}: WidgetVA: A Widget-Centric Framework and Benchmark for Agentic Visual Analytics}}

%\IEEEaftertitletext{%
%    \vspace{-0.5\baselineskip}
%    \begin{minipage}{\textwidth}
%        \centering
%        \includegraphics[width=0.96\textwidth,keepaspectratio]{figs/teaser.new_cropped.pdf}
        % \captionof{figure}{Overview of \name, a widget-centric agentic visual analytics framework. The framework provides interaction capabilities, supports both integrated VA systems and existing VA pages, and organizes multi-step analysis through reusable workflows. \bench further enables diagnostic evaluation of agent behavior across single- and multi-widget tasks.}
%        \captionof{figure}{In \revise{\name}, interactive views are exposed through a unified widget so that humans and agents can explore and analyze together in the same workspace. \revise{\name supports wrapping an existing VA page to make it agent-operable or composing a new system from widgets, and it provides reusable workflows to support multi-step planning. We also construct \bench to evaluate agent performance and exposes persistent limitations for future work.}}
%        \label{fig:teaser}
%    \end{minipage}
%    \vspace{0.5\baselineskip}}

\maketitle

\begin{abstract}
Visual analytics (VA) enables sensemaking through interactive visualization, but effective analysis often requires experts to translate high-level intents into long sequences of interface operations and iteratively interpret visual feedback.
We study whether modern vision-language models (VLMs) can take on this role as autonomous VA operators that observe the interface, plan multi-step exploration, execute interactions, and adapt based on intermediate visual feedback.
To support systematic development and evaluation, we first introduce \revise{\name}, a widget-centric agentic VA framework that standardizes interactive components as structured widgets with unified action (\eg, filter and zoom) and perception-query (\eg, selection summaries) APIs.
\revise{This standardization supports two modes of system construction: wrapping an existing VA system to make it agent-operable without rebuilding it, and composing a new system from widgets as modular building blocks.
To help agents coordinate across widgets rather than plan each interaction from scratch, each widget further packages reusable analytical workflows, giving agents more than a bare set of callable functions to plan over.}
Building on this framework, we present \revise{\bench}, a benchmark of \revise{single- and multi-widget} VA tasks that require agents to perform multi-step interactions to uncover evidence and produce verifiable results.
\revise{Each task also provides fine-grained reference annotations so that \bench can score Answer, Reference Trace Similarity, and State separately rather than collapsing agent performance into one success score.} % , including an expected answer, an executable reference trajectory, and task-relevant state conditions,
\revise{Experiments across multiple VLMs show that our framework provides an effective scaffold for agentic VA, while the diagnostic measures expose persistent limitations for future work. }
\revise{
The \name framework and \bench have been released in   \href{https://github.com/Hiverwin/widgetva}{https://github.com/Hiverwin/widgetva}.
}

\end{abstract}


\section{Introduction}
\label{sec:introduction}

\begin{figure*}[t]
  \centering
  \includegraphics[width=\textwidth]{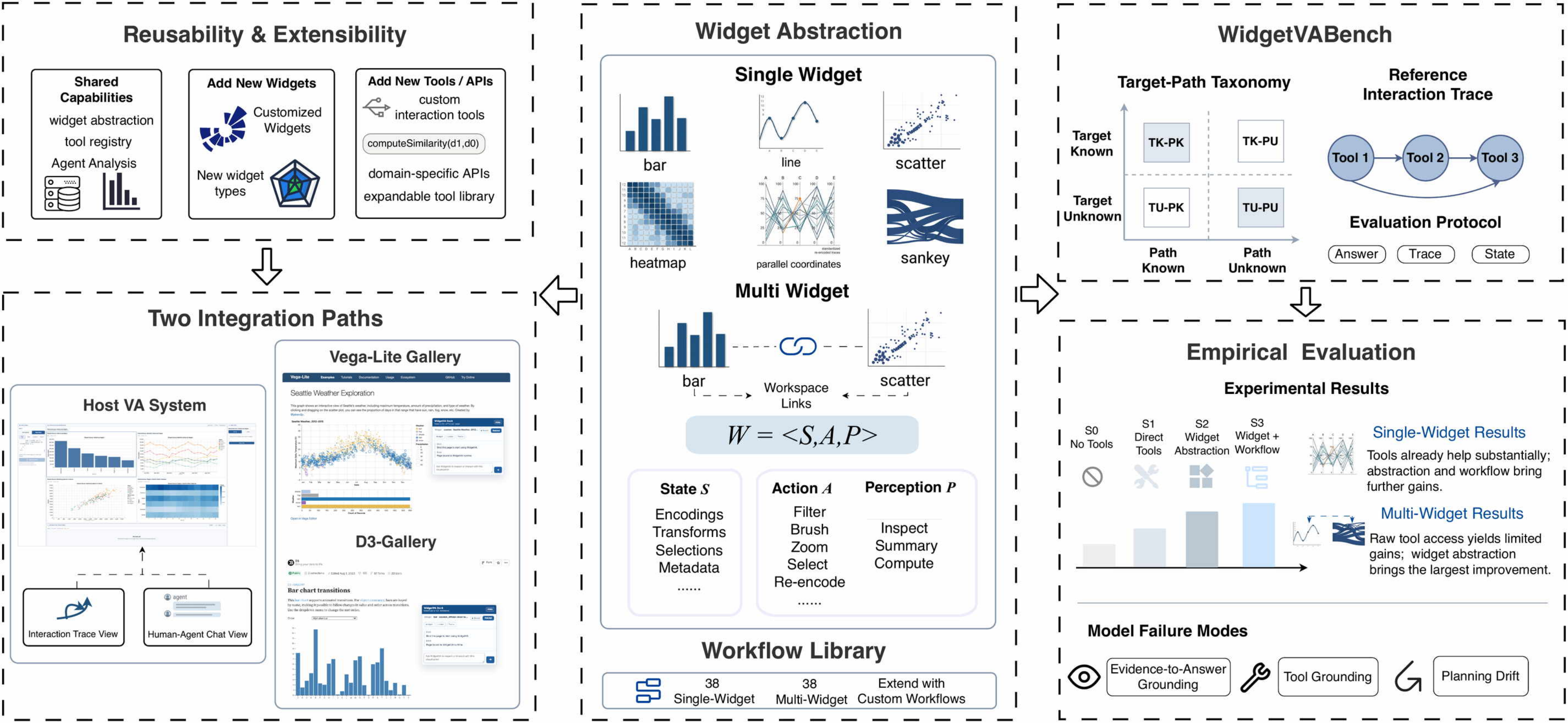}
  \captionof{figure}{In \revise{\name}, interactive views are exposed through a unified widget so that humans and agents can explore and analyze together in the same workspace. \revise{\name supports wrapping an existing VA page to make it agent-operable or composing a new system from widgets, and it provides reusable workflows to support multi-step planning. We also construct \bench to evaluate agent performance and exposes persistent limitations for future work.}}
\label{fig:teaser}
\end{figure*}

\IEEEPARstart{V}{isual} analytics (VA) combines interactive visualization with computational analysis to support sensemaking over complex data~\cite{keim2008visual,liu2017towards,Liu2025}.
However, effective VA requires analysts to iteratively formulate hypotheses, translate \revise{high-level intents} into sequences of interface operations (\eg, filtering, brushing, zooming), interpret \revise{intermediate} visual feedback, and revise their analytical strategy.
This iterative loop is cognitively demanding and requires expertise in both the application domain and the VA system itself~\cite{yuan2021survey,yang2024foundation}.
As a result, it limits who can effectively use VA systems and how quickly insights can be reached.

% \revise{This burden cannot be removed simply by letting an agent analyze raw data through code.
% Direct data analysis is appropriate for many purely computational questions, but it bypasses the exploration tools that VA already provides, such as filtering, brushing, zooming, re-encoding, and coordinated views.
% These interactions structure search and make intermediate patterns visible in a form that a human analyst can inspect.
% If an agent instead operates the same VA workspace, the human and the agent can share, revise, and continue from the same views and interaction state, rather than reconciling a separate code-based analysis with the visual interface.
% Because each step leaves an inspectable visual and state trace, analysts can follow how a conclusion was reached and decide whether to accept, correct, or override it.
% Our aim is therefore not to replace raw-data analysis, but to make interactive VA environments agent-operable for human--agent co-analysis.}

\revise{Recent progress in large language models (LLMs) and vision-language models (VLMs) has created an opportunity to delegate parts of this interaction loop to an agent.}
\revise{Letting an agent analyze raw data through code is one way to reduce this burden for many purely computational questions, but it bypasses the exploration tools that VA already provides, which structure search and leave an inspectable trace for a human analyst to follow, correct, or build on.
Rather than bypass these tools, a growing number of recent systems augment themselves with} large models to facilitate tool onboarding~\cite{Zhao2024LEVA}, data exploration~\cite{Zhao2024LightVA, Zhao2025ProactiveVA}, and automated insight summarization~\cite{Wang2025ChartInsighter}.
However, models in these systems are largely positioned as assistants that do not autonomously operate VA systems~\cite{Amershi2019Guidelines}, \revise{and their reasoning remains tightly coupled to application-specific states, functions, and front-end operations, which limits reuse across VA systems.
On the other hand, general-purpose GUI agents provide broader application coverage, but screenshots and low-level mouse or keyboard actions expose limited analytical semantics for planning and state verification~\cite{Cheng2024SeeClick,Niu2024ScreenAgent,Xie2024OSWorld}.
Neither route yet provides a systematic, reusable way to make VA interfaces agent-operable across systems, or evidence of how far current VLMs can already carry out such visual analysis tasks.
Our goal is therefore to build the former and measure the latter: to make interactive VA environments agent-operable for human--agent co-analysis, where agents share, revise, and continue analysis from the same views and interaction state as a human.}
% \revise{Our goal is therefore to build the former and measure the latter: to make interactive VA environments agent-operable so that a human and an agent can explore through the same views, operations, and interaction state.}
% \revise{Instrumenting VA in this way gives them a shared, inspectable interaction language for co-analysis, rather than leaving the agent to return a disconnected code-based answer.}}

\revise{Realizing these two goals, however, raises four linked technical challenges.
First, different VA implementations expose incompatible, provider-specific states and operations, so it is unclear how to define a single agent-facing representation that captures the same analytical semantics, such as encodings, transforms, and selections, without depending on any one rendering library.
Second, when multiple widgets are coordinated, such as a selection in one view filtering another, the resulting state changes must remain attributable to the action that caused them, rather than appearing as untraceable side effects.
Third, translating a high-level analytical goal into a correct sequence of concrete operations requires the agent to verify each intermediate outcome and adapt its plan accordingly, rather than committing to an unverified action sequence end-to-end.
Finally, final-answer accuracy alone cannot reveal whether a failure arises from answer synthesis, operation grounding, or workspace-state construction, yet scoring against one fixed reference trace would unfairly penalize an agent that reaches an equally valid state through a different path.}
% Although recent benchmarks incorporated interactive dashboards~\cite{Kartha2026DashboardQA,Wu2026VizAgentBench}, diagnosing an agent's behavior along these lines still requires explicit, decomposed representations of its operations and resulting semantic states.

% Original: To address these challenges, we first propose a widget-centric agentic VA framework that standardizes interactive visualization components into a structured, model-friendly abstraction (\Cref{fig:teaser}).
To address these challenges, we introduce a widget-centric agentic VA framework that \revise{separates a model-facing interaction abstraction from the host system's visualization grammar and rendering implementation (\Cref{fig:teaser}).}
Each widget is represented as $W_i=(S_i,\mathcal{A}_i,\mathcal{P}_i)$, comprising its semantic state, typed action capabilities, and side-effect-free perception queries.
\revise{Individual widgets compose into a workspace representation that records shared state, semantic links, and coordination effects across views.
% Original: This representation supports both adapting existing VA systems to be agent-operable without rebuilding them from scratch, and constructing new VA systems by composing widgets as reusable building blocks.
\revise{The same contract is therefore usable in two ways: wrapping an existing VA system so that it becomes agent-operable without being rebuilt, and composing widgets to assemble a task-specific VA workspace quickly.}
Operating over this workspace, a closed-loop scaffold repeatedly observes the current widget or workspace state, selects and executes an action or perception query, incorporates the returned evidence, and plans the next step.
To facilitate agent planning, an analytical workflow library supplies reusable analysis workflows for translating recurring visual-analysis goals into widget-level action and perception sequences.
}

% Original: On top of this abstraction, we provide a concrete implementation of the framework and develop \bench, a benchmark of representative VA tasks where interaction is necessary to surface evidence and reach verifiable outcomes.
% Original: On top of this abstraction, we provide a concrete implementation of the framework and develop \bench, a benchmark of representative VA tasks with graded interaction depth, reference tool traces, and verifiable results aligned with annotated widget trajectories.
% Original: Alongside final answers, we provide fine-grained evaluation signals such as intermediate view state, tool-call traces, and rationale annotations to enable  behavioral analysis.
Built on our framework, we develop \revise{\bench}, a benchmark of representative VA tasks where interaction is necessary to surface evidence and reach verifiable results.
\revise{Each task packages an expected answer, an executable reference trace, and task-relevant conditions on the final semantic state.}
\revise{Tasks are further organized by whether their analytical target and path are known, following the knowledge-gap characterization of analytic guidance~\cite{ceneda2017characterizingGuidance}.
Together, these reference annotations and the Target--Path task taxonomy support fine-grained diagnosis: rather than reducing agentic VA to a single end-task score, \bench scores Answer, Reference Trace Similarity, and State as separate dimensions and further breaks results down by knowledge-gap category.}
These signals help diagnose failures in answer reasoning, operation grounding, and analytical-state construction, respectively.
\revise{The resulting failure patterns can inform the next generation of agent-ready VA systems, for example by indicating whether action APIs, exposed state, or verification feedback should be redesigned.}
\revise{Moreover, since human and agent operations already share the same contract, ordinary analysis sessions can be recorded as structured traces for later data collection, skill construction, or agent training.}

% Original: We evaluate state-of-the-art VLMs on \bench to characterize a capability frontier for agentic VA.
% Original: We evaluate multiple state-of-the-art VLMs across execution settings that vary access to interaction tools and structured scaffold support.

% Original: In addition, our analysis surfaces recurring failure modes, such as premature action commitment, insufficient state tracking, and brittle error recovery, which limit today's models as autonomous VA operators and motivate future work.
% Original: The current results show that interaction improves Answer scores over the no-tool condition and that the structured scaffold further improves the tool-only comparison across the reported task strata.
% The largest gain comes from tool access itself, which is expected once a model can operate the interface at all.
To assess the framework's contribution, we compare multiple state-of-the-art VLMs \revise{across four settings: No Tools, Direct Tools, Widget Abstraction, and Widget Abstraction + Workflow.
The results show consistent gains over the baselines across all three VLMs, indicating that the framework is effective at making VA interfaces agent-operable.}
\revise{Our diagnostic analysis further identifies recurring errors in chart perception, tool grounding, and multi-step planning that persist even under the framework.}
We use these findings to characterize current limitations of agent--VA interaction \revise{rather than to present a durable ranking of rapidly evolving models.}

% Original: To summarize, this paper makes three contributions .
In summary, this paper makes three contributions:
\begin{itemize}[]
% Original: A widget-centric agentic VA abstraction that standardizes interactive components and exposes unified APIs for actions and structured perception queries.
\item \revise{A scoped, model-facing widget and workspace abstraction, together with a concrete implementation that makes interactive VA components agent-operable and their state transitions inspectable.}
% Original: \bench, a benchmark of long-horizon, interaction-centric VA tasks with explicit success criteria to enable reproducible evaluation and comparison.
\item \revise{\bench}, an interaction-centric benchmark \revise{spanning single-widget and coordinated multi-widget VA systems, with Target--Path task categories and separate annotations for answers, reference traces, and final states}.
% Original: An empirical evaluation and analysis of multiple VLMs on \bench, identifying capability limits and failure patterns that inform both model development and VA system design.
\item A diagnostic evaluation of multiple VLMs that \revise{compares levels of interaction support} and identifies limitations in perception, operation grounding, \revise{state construction}, and multi-step planning.
\end{itemize}

\section{Related Work}
\label{sec:related_work}

\subsection{\revise{Agentic Visual Analytics and Model-Facing Interfaces}}

The study of conversational visualization has long explored the combination of natural language and visualization for analysis (\eg, when and how to embed visual context in conversation)~\cite{Hearst2019ChartWithThat}.
Recent breakthroughs in large language models (LLMs) and vision-language models (VLMs) have pushed this area forward, and VIS/TVCG research has shown that LLMs can be integrated into VA systems to
support onboarding, exploration, and summarization. For instance, LEVA interprets visualization designs and view relationships from system specifications, suggests insights based on system status, and assists
in report generation based on interaction history~\cite{Zhao2024LEVA}. To complement this, ProactiveVA points out that even the latest LLM-assisted VA systems are only useful when explicitly asked for assistance, and proposes an LLM-assisted UI agent that monitors interaction logs to proactively identify when the user needs assistance, reason about what to provide, and execute that assistance through the interface in an interpretable and controllable manner~\cite{Zhao2025ProactiveVA}.
Other related work also investigates how to improve context and intermediate insight management in LLM-assisted analysis systems (\eg, InsightLens) or integrate text and visualization for interactive data analysis (\eg, DASH)~\cite{Weng2025InsightLens,Bromley2024DASH}.
Overall, these works indicate a clear trend: VA systems are moving from passive systems to increasingly agentic systems.
However, most of these systems are tightly coupled to a specific front-end and interaction design, which makes it hard to reuse the same model in different tools or to compare agent behaviors.

\revise{Other work has therefore sought more general and reusable methods that are not tied to a single VA front end.}
\revise{A visualization grammar such as Vega-Lite provides a portable specification of how data, visual encodings, and interactions are compiled into views~\cite{Satyanarayan2017VegaLite}, but such a grammar alone does not provide agentic capabilities for inspecting and operating an instantiated visualization.}
Multimodal UI agents instead pursue generality by operating directly on screenshots with low-level actions.
SeeClick and ScreenAgent improve GUI grounding and computer control by learning to identify and manipulate GUI elements from visual input~\cite{Cheng2024SeeClick,Niu2024ScreenAgent,Xie2024OSWorld}.
These examples illustrate the potential of computer-using VLMs, but pixel-level interfaces are often opaque about intermediate state and debugging, particularly for analytical tasks that demand precise state monitoring, verification, and user control.
\revise{Our widget abstraction therefore sits between these two directions: it reuses instantiated visualization components rather than replacing their host grammar, and exposes them through structured states, intent-level actions, and perception queries with consistent inputs and outputs.}

\subsection{Benchmarks for Visualization Reasoning}
Current visualization benchmarks and studies are largely based on static charts, focusing on visual question answering and visualization literacy.
For example, Xu and Wall investigate LLMs' capability to execute low-level analytic tasks directly on SVG visualizations of charts~\cite{Xu2024LLMTasksSVG}.
However, practical dashboards involve complex multi-view visualizations and rich interactions, where important evidence is often hidden until filtered, drilled down, or hovered over.
DashboardQA brings agentic question answering to real-world Tableau dashboards in a virtual desktop environment, requiring VLM GUI agents to execute multi-step GUI tasks while preserving intermediate state~\cite{Kartha2025DashboardQA}.
VizAgentBench also assesses multimodal agents on real-world, interactive, and coordinated dashboards with a declarative interaction API that decouples perception and action~\cite{Wu2026VizAgentBench}.
These benchmarks have greatly contributed to the area by recognizing interactivity as a fundamental aspect of dashboard reasoning.
However, the state of evaluation is still fragmented: existing benchmarks are QA-focused and mostly treat the interface as a pixel-level domain, which makes it hard to analyze agent failures: whether they arise from visual perception, fragile state management, poor long-horizon planning, or inadequate error recovery.
Moreover, they offer little leverage for a systematic investigation of the role of interface representations and feedback mechanisms, \ie, what the system reveals to the agent beyond the raw pixel data, in facilitating robust autonomous operation, which is a problem that directly relates to VA system design.
To address this problem, \revise{\bench} emphasizes tool-mediated interaction with visualization widgets, provides both objective QA and exploratory questions, and records structured intermediate \revise{tool traces and final states} to facilitate fine-grained analysis of agent behavior.

\subsection{\revise{Reusable Analytical Workflows}}
\revise{Visualization research has developed task and interaction taxonomies to describe recurring analytical goals. Amar~\etal identify and classify low-level activities as retrieving values, filtering, finding extrema, characterizing distributions, detecting anomalies, and examining correlations~\cite{Amar2005LowLevel}. Brehmer and Munzner connect an analyst's motivation, target, and method through a multi-level task typology~\cite{Brehmer2013Typology}, while interaction taxonomies describe intents~\cite{Yi2007Interaction,Heer2012InteractiveDynamics}. These abstractions establish the analytical operations that a VA system should support. They provide less guidance on how several operations should be ordered, when an intermediate result should be inspected, or how evidence should be transferred between coordinated views during an agent's execution.}

\revise{
Prior work has addressed parts of this problem. Provenance models capture how an analysis develops over time~\cite{Gotz2009Provenance}, and grammar-based methods map low-level interaction logs to higher-level analytical activities~\cite{Gathani2022Grammar}. Gadhave~\etal represent interactions at a semantic level so that completed analyses can be reused with updated data~\cite{Gadhave2022ReusableWorkflows}. These approaches are effective for preserving and reapplying previously demonstrated procedures, but generally assume that a suitable analysis has already been carried out and recorded. ATWL instead represents workflows at a higher level as typed artifacts connected by intent-bearing transformations, supporting their description, comparison, and reuse across analytical domains~\cite{Andrienko2026ATWL}. Such representations characterize the structure of analytical processes, while offering limited guidance on which recurring analysis pattern an agent should adopt for a new visual-analysis query.}

\revise{
Our workflow library addresses this gap by summarizing common visual-analysis patterns as planner-facing knowledge. It covers patterns within single views and across multiple views, drawing on established analytical-task and coordinated-view practices~\cite{Baldonado2000MultipleViews,Roberts2007CMV}. Making this library available to the agent's planner reduces the need to derive an analysis strategy from scratch and helps agents complete a broad range of classical visual-analysis tasks more consistently.}
\section{Widget-Centric Agentic VA Framework}
\label{sec:framework}

We aim to develop and evaluate VLM agents that can complete VA tasks by operating interactive interfaces.
However, current VA tools expose heterogeneous interfaces that make agent behavior hard to reproduce and
diagnose. A standardized interface is therefore necessary to make VA
systems easier for agents to operate, as well as to make agent behavior
reproducible and diagnosable. To this end, we present \revise{\name}, a widget-centric
framework that standardizes interactive visualization components as
widgets with explicit state, action APIs, and perception APIs.
%Thisdesign enables VLMs to operate VA interfaces through standardizedAPIs and supports closed-loop interaction and systematic evaluation

\subsection{Design Requirements}
\label{sec:framework_rationale}
% \revise{I didn't modify this subsection}

The framework was iteratively refined through biweekly interviews with three experts (E1--E3) in developing visual analytics systems and agent systems.
E1 is a visualization researcher with extensive experience in building VA systems.
E2 and E3 are Ph.D. students in computer science with one year of experience building tool-using agents.
None of them are coauthors of this paper.
We distilled their feedback from these discussions into three design requirements.
% Original: We distilled their feedback into three design requirements. % , each paired with a concrete design choice in our widget-centric framework

% Original: When operating real interfaces, agents that rely on screenshots and coordinate-based actions often incur long action sequences and face unreliable grounding.
\textbf{R1: Intent-aligned, structured interaction.}
The supported interactions should be both model-friendly and analytically meaningful.
% Original: Rather than requiring pixel-level control (\eg, mouse clicks and keyboard shortcuts), the high-level interactions that match analysts' intent (e.g., filter, zoom, select, and re-encode) are more meaningful and easier to reason about for both humans and agents.
Rather than requiring pixel-level control (\eg, mouse clicks and keyboard shortcuts), the framework exposes high-level interactions that match analysts' intent (\eg, filter, zoom, select, and re-encode), which are more meaningful and easier to reason about for both humans and agents.
E2 commented, \textit{``a model-friendly interface should expose intent-level operations with typed parameters, so an agent can plan and act in data space rather than chase pixels.''}
E3 also argued that while pixel-level control seems more robust, it may fail with unreliable grounding and introduce unnecessary complexity for the model.
Moreover, this design reduces ambiguity for the model and yields cleaner, comparable traces.
The trace of interaction history, including tool calls and parameters, can be directly inspected and evaluated across runs.
% We meet this requirement by standardizing interactive components as widgets and exposing their operations as semantic tool calls with explicit parameters~\cite{AXIS2025}.

% Original: Interactive analysis requires more than executing actions; it requires verifying what changed and extracting evidence to decide the next step.
% Original: Pixel-only observation makes such verification hard, especially for dense marks or tooltip-dependent values.
% Original: \paragraph{R2: Enable closed-loop evidence acquisition and verification.}
\textbf{R2: Queryable intermediate states.}
VA tasks are typically performed through a sequence of interactions rather than a one-shot problem~\cite{yang2020diagnosing,yang2021interactive}.
After each operation, users need to verify what changed, extract evidence, and decide what to do next.
E2 emphasized: \textit{``The agent must be able to immediately read back outcomes after taking an action. Otherwise, the agent cannot reliably recover from mistakes or refine hypotheses.''}
E1 added that it might be useful to provide the results in both text and visual format:
\textit{``while the text format is more precise, the visual format is more intuitive and easier to understand for some tasks, such as identifying clusters in a scatterplot.''}
Therefore, the framework is designed to pass both the updated state (text format) and the new rendered view (visual format) into the next interaction run.
In addition, this design supports a more robust evaluation by comparing the interface's intermediate states rather than relying solely on the final answer and the trace of the interaction history.

\textbf{R3: Portable and unified interface.}
A portable and unified interface is necessary to abstract heterogeneous VA systems into consistent, model-friendly schemas and tool signatures.
Such uniformity reduces the learning and control burden for VLM agents.
E1 commented, \textit{``By making each interactive visualization component a plug-and-play object with a standardized API, developers can easily build VA systems and directly reuse the agent capabilities inherited from the framework.''}
Meanwhile, this design makes the framework extensible: new visualizations can be easily supported by implementing the widget schema.

% To meet the three disgn requirements, we propose the following framework (\Cref{fig:framework}).
% \begin{figure*}[t]
%   \centering
%   \includegraphics[width=\textwidth]{figs/framework.pdf}
%   \caption{
% Overview of our widget-centric agentic VA framework.
% Interactive charts are abstracted as widgets with standardized state, action, and perception interfaces, allowing tool-using agents to operate heterogeneous VA systems through a common semantic contract.
% The same abstraction supports reuse and extension across new tasks, widgets, and front-end implementations.
% }
%   \label{fig:framework}
% \end{figure*}

\subsection{Widget Abstraction}
\label{sec:framework_widget}
% \revise{This part remains the original structures, explaining state,action and perception but add more details}
% Original: Guided by the design requirements, we designed a widget as an agent-facing abstraction of an interactive visualization component.
% Original: Each widget exposes its interaction-relevant analytical context and capabilities through an implementation-agnostic interface.
% Original: Specifically, each widget $\mathcal{W}$ consists of \textbf{widget state} $S$, \textbf{action} $\mathcal{A}$, and \textbf{perception queries} $\mathcal{P}$.
Guided by the design requirements, we take a widget as the unit of agent-facing abstraction.
\revise{A widget is self-contained enough to expose a coherent analytical state and intent-level operations (R1--R2), and compositional enough to serve as a building block for coordinated workspaces (R3).}
\revise{Each widget therefore exposes its interaction-relevant context and capabilities through an implementation-agnostic contract} $\mathcal{W}=(S,\mathcal{A},\mathcal{P})$, comprising \textbf{widget state} $S$, \textbf{action} $\mathcal{A}$, and \textbf{perception queries} $\mathcal{P}$.

% \revise{The contract exposes the analytical context needed for agent interaction through an implementation-agnostic interface.}
%A system built with Vega or Vega-Lite~\cite{Satyanarayan2017VegaLite}, for example, may retain its native specification internally; systems based on other libraries need not translate their views into that grammar.
%In either case, an adapter projects the interaction-relevant semantics of the host view into the same widget abstraction.

% Each widget $W_i$ is represented as
% \begin{equation}
% W_i = (S_i, \mathcal{A}_i, \mathcal{P}_i),
% \end{equation}
% where $S_i$ captures its \revise{current analytical} state, $\mathcal{A}_i$ represents the actions which changes state, and $\mathcal{P}_i$ represents the perception queries \revise{which can be used to obtain evidence.}
% \revise{These elements form an interaction contract: actions express analytical intent (R1), state and perception make intermediate outcomes accessible (R2), and adapters preserve these semantics across heterogeneous implementations (R3).}

\subsubsection{Widget State}

The widget state specifies the widget's view and interaction configuration, enabling a visualization to be rendered from the state.
% This state allows an agent to interpret what the widget shows and allows tools to update the view reproducibly.
Inspired by the visualization pipeline proposed by Card~\etal~\cite{card1999readings}, \revise{we represent the state of a widget as}
\revise{
\begin{equation}
  S_i = (D_i, E_i, T_i, V_i, \mathrm{Sel}_i, F_i),
  \end{equation}
  with the following components:}
  \begin{itemize}
  \item \revise{\textbf{Data Context ($D_i$)} records references to the source and important information about the current data view, such as the numbers of visible and selected records. }
  \item \revise{\textbf{Encoding Context ($E_i$)}} \revise{records the semantic bindings between data fields and visual channels} such as position (x, y), color, size, and shape.
  \item \revise{\textbf{Transformation Context ($T_i$)}} records transformations, such as filtering, sorting, and aggregation, \revise{as well as their source references when a transformation is induced by an action issued on a linked widget.}
  \item \revise{\textbf{View Context ($V_i$)}} specifies view-level configuration and transforms, such as scale domains and ranges, axis configuration, and pan/zoom extents when applicable.
  \item \revise{\textbf{Selection Context ($\mathrm{Sel}_i$)} records active point, interval, or predicate selections.}
  \revise{\item \textbf{Interaction Feedback ($F_i$)} records transient evidence returned by the widget, such as hovered items and tooltips.}
\end{itemize}

% \revise{This definition separates the analytical state from the underlying implementation-specific view representation, e.g., Vega/D3/ECharts.}
% For example, a Vega-Lite specification may be retained by a Vega-Lite adapter to render and update a view, while a D3 or ECharts adapter can expose the same categories of state without translating the whole visualization into Vega-Lite.
% As a result, two visually different implementations can present a comparable agent-facing state as long as they support the same analytical semantics.}

% \revise{In practice, a VA system usually contains multiple coordinated widgets.
% We therefore compose individual widget states into a workspace state
% \begin{equation}
% S^{\mathrm{ws}} = (\{S_i\}_{i=1}^{n}, S^{\mathrm{shared}}, L, C),
% \end{equation}
% where $S^{\mathrm{shared}}$ stores shared widgets state context, $L$ describes the links and their topology among widgets, and $C$ records the resulting coordination relations.
% This workspace-level state makes multi-widget interactions explicit.
% For example, a selection in one widget can become a filter or synchronized domain effect in another widget, while the resulting states remain attributable to the source action and link.
% Each committed change is assigned a state identifier and can be recorded as a snapshot, which supports replay and provenance without requiring the agent to reconstruct the prior interface from screenshots. }

\revise{In practice, a VA system usually contains multiple coordinated widgets, so we support composing individual widget states into a workspace state that captures shared state across widgets, their link topology, and the coordinated interactions.}

\subsubsection{Action}
\label{sec:action-api}
Each action corresponds to an intent-level interaction (\eg, filter, zoom), which takes the current widget state and typed parameters as input and returns an updated widget state.
Rather than enumerating all concrete actions per widget, \revise{we organize them by their underlying intent, drawn from interaction techniques commonly supported by interactive visualizations, into six primitives}: data transformation (\textsc{Filter}, \textsc{Sort}, \textsc{Drill-down}, \textsc{Aggregate}), visual mapping (\textsc{Re-encode}), view transformation (\textsc{Zoom}), interactive selection (\textsc{Brush/Select}, \textsc{Highlight}, \textsc{Focus}), analytic augmentation (\textsc{Overlay/Annotate}), and navigation (\textsc{Navigate}, \textsc{Add/Remove}).
These intents are instantiated as widget-specific tools.
\revise{For example, the \textsc{Zoom} primitive becomes \texttt{line.zoomXRegion} for a line chart and \texttt{scatter.zoomDomain} for a scatterplot.}
Not every widget supports every primitive, and some widgets may provide additional actions beyond existing primitives (\eg, expanding stacked bars).

\subsubsection{Perception Queries}
\label{sec:perception-api}
Perception queries take the current widget state as input and return structured, machine-readable information about what the widget currently shows.
Unlike actions, perception queries are side-effect-free, \ie, they deterministically return the same result given the same state and parameters. 
This separation enables closed-loop exploration, where the agent alternates between executing actions and issuing perception queries to validate intermediate results.
Similar to actions, we derive perception primitives and organize them into three categories: \textsc{Inspect} exposes structured evidence such as view configuration and raw data, \textsc{Summarize} returns lightweight aggregates for quick comparison and sanity checks, and \textsc{Compute} derives higher-level analytical results such as correlations.

%\subsubsection{Perception}
%\label{sec:perception-api}
%Perception queries take the current widget state as input and return structured, machine-readable information about what the widget currently shows.
%Each widget exposes perception queries according to its widget kind. \revise{Their descriptors illustrates what question a query can answer and what context it operates on. The agent can therefore combine the
%current state and the user query to decide which perception query is needed for evidence. After executing the perception query, agent can receive analytical evidence from a referenced data view or selection, or use the information receiverd to 
%check whether the executed action produces the intended change.
%With the perception queries, the agent can therefore reason jointly over the updated view, the widget state, and the 
%returned perception evidence at each step, satisfying the requirement that analysis remain grounded in the current visual and interaction context (R1).}

% \revise{Together, widget state, action, and perception form a common interaction contract between the agent and the VA workspace.
% The state illustrates the analytical context in which an interaction occurs; actions modify the state, while perceptions derive evidence from it. }
\revise{Together, they form a common interaction contract between agents and VA systems.
By decoupling this contract from implementation-specific view representations (\eg, Vega, D3, ECharts), agents gain a unified interface for perceiving and manipulating widgets across heterogeneous toolkits.}

\subsection{Orchestrating Agent-Widget Interaction}
\label{sec:agentic_orchestration} 

\paragraph{Interaction Orchestration}
\revise{
The widget abstraction describes the analytical state and the capabilities available during analysis.} However, visual analysis usually requires multiple interactions: an action changes the state, while a perception query inspects
the resulting view or obtains further evidence needed for subsequent analysis. \revise{We therefore introduce a lightweight orchestration layer to coordinate the use of the widget contract across successive agent turns.} At each turn, the orchestrator provides the agent with the user query, the current rendered view, the corresponding widget \revise{or workspace} state, and \revise{the available capabilities}. The agent may answer directly or invoke an action or perception capability based on this context. \revise{After the call is executed, the updated state or returned evidence is incorporated into the next turn. In coordinated views, changes propagated through widget links are also reflected in the workspace state and are visible to the agent at each turn.} Consequently, each subsequent plan is grounded in the actual outcome of the preceding interactions.

\paragraph{Analytical Workflow Library}
\label{sec:bench_workflows}

\revise{
During the development of our work, we repeatedly observed common sequences for recurring analytical goals, both in agent runs and in human annotation.
These patterns were useful because they turned a multi-step analysis into a reusable procedure, rather than requiring the agent to reconstruct a plan from a bare tool list at every turn.
We therefore compiled a workflow library from two complementary sources.
First, we reviewed recurring task abstractions and interaction intents reported in visualization task and interaction taxonomies~\cite{Amar2005LowLevel,Yi2007Interaction,Brehmer2013Typology}.
%Shneiderman1996
Second, we examined human-annotated interaction traces collected during task generation and identified common patterns that repeatedly appeared across annotated instances.
We then summarized these general patterns and common analytical procedures to create a workflow library.}
\revise{
The resulting library contains \textbf{38} single-widget workflows and \textbf{38} multi-widget workflows spanning two to six widgets.
Each workflow records its relevant widget types, analytical intent, representative scenarios, and guidance on how to perform the analysis.
The single-widget workflows include procedures such as constrained ranking and interval-specific anomaly search, while the multi-widget workflows cover procedures such as aggregate-to-detail investigation and repeated cohort comparison.}

\begin{figure*}[t]
  \centering
  \includegraphics[width=\textwidth]{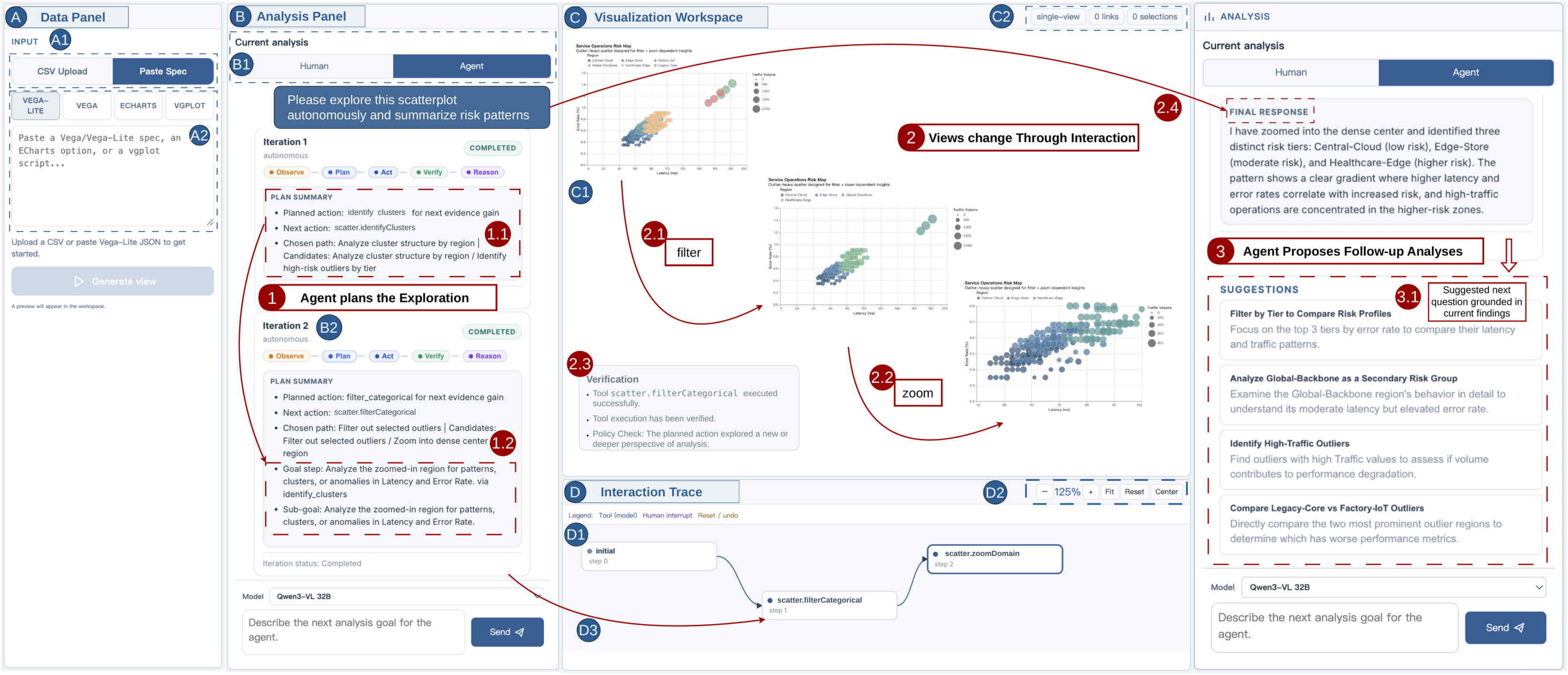}
  \caption{Overview of our \revise{integrated analysis environment}.
(A) The \textbf{Data Panel} \revise{initializes an analysis session from uploaded tabular data or a supported visualization specification.}
(B) The \textbf{Analysis Panel} \revise{supports mixed-initiative analysis by allowing users to record useful insights and inspect the agent's iterative analysis process.}
(C) The \textbf{Visualization \revise{Workspace}} \revise{renders the current visualization state.}
(D) The \textbf{Interaction Trace Panel} records the analysis \revise{as a branch-aware provenance graph and links each analysis step to its corresponding iteration card and workspace state.}}
\label{fig:system_overview}
\end{figure*}

%% ---------- 3.4  Implementation ----------
\subsection{Implementation}
\label{sec:framework_implementation}
% \revise{rewrit this part totally, Widget Kit subsection illustrates the architechture and how to extend the action and perception capability; integration explains the implementation in 2 methods}

\revise{We implement the proposed framework as \textit{WidgetVA Kit}, a JavaScript toolkit that realizes the widget abstraction and exposes it for integration with practical VA environments.
It supports both adapting existing VA systems to be agent-operable without rebuilding them from scratch, and constructing new VA systems by composing widgets as reusable building blocks.
% This section describes how the widget abstraction is organized and how can it be extended, and then illustrates two ways of using the abstraction in practice: integrating WidgetVA into a VA system, or attaching it to an existing VA page through a browser extension.
}

% This representation supports both adapting existing
% VA systems to be agent-operable without rebuilding them
% from scratch, and constructing new VA systems by com-
% posing widgets as reusable building blocks.

\subsubsection{\revise{WidgetVA Kit}}
\paragraph{\revise{Architecture}}
\revise{The implementation follows the same widget-centric abstraction introduced in \Cref{sec:framework_widget}.
The kit defines shared contracts for widget and workspace states, actions, perceptions, and inter-widget links, so that these elements have a consistent representation across different visualization implementations.
For each visualization kind, a \textit{widget family} instantiates this contract by defining the actions and perception queries supported by that kind, together with the interaction information needed to operate the widget.}

\revise{A provider denotes the underlying visualization implementation used to realize a widget, such as Vega-Lite, D3, ECharts, or VGPlot.
The corresponding provider adapter connects the standardized widget semantics to the provider's native view and interaction mechanisms.
It applies widget-state changes to the rendered view and captures provider-specific human interactions so that their effects are reflected back in the widget state.
This separation allows the analysis workflow to operate on a consistent widget interface while each visualization retains its native rendering and interaction logic.
Multiple widget instances can further be composed through the workspace representation, enabling the same abstraction to support both individual visualizations and coordinated multi-view systems.}

\paragraph{Extensibility}
Our implementation currently supports six widget types: bar charts, line charts, scatterplots, heatmaps, parallel coordinates, and Sankey diagrams.
\revise{We include parallel coordinates and Sankey diagrams to demonstrate that the widget abstraction can accommodate more complex interactive visualizations beyond conventional statistical charts.}
Developers can extend existing widget types with a new action or perception query by providing the descriptor that \revise{specifies its name, applicable widget kinds,} typed parameters, \revise{and expected effects}.
Developers can also register a new widget type \revise{by providing a widget-family module that declares the widget kind, supported actions, perception queries, and executable implementations}.
%Once registered, the new widget is exposed to the planner through the same runtime interface as built-in widgets, without modifying the agent loop.
%\vica{Introduce how to do it}
% Developers can add a new interaction to an existing \revise{widget type by adding new actions or perception queries to its widget contract.}
% The developer provides \revise{an action descriptor that specifies its name, applicable widget kinds,} typed parameters, \revise{and expected state effects}.
% A perception query is extended in the same way: \revise{its descriptor specifies the query interface and expected output.}
% Once registered with the widget family, the new interaction is exposed together with the built-in capabilities and can be selected and invoked through the same analysis process.

\subsubsection{\revise{Integration}}
\revise{We demonstrate two ways to integrate WidgetVA with practical VA environments.
% Once a visualization is exposed through the widget abstraction, WidgetVA can be used through two integration methods.
% For a VA system, developers can integrate the WidgetVA Kit directly into its visualization and interaction components.
% For an existing VA page, developers can attach the WidgetVA Kit to the page and expose the same widget capabilities through a browser extension.
}
\paragraph{\revise{Integration with Existing VA Systems}}
\revise{For existing VA systems, such as interactive multi-view examples in the Vega-Lite gallery\footnote{https://vega.github.io/vega-lite/examples/interactive\_seattle\_weather.html}, we provide lightweight integration through a browser extension.
The extension connects to the page's native visualizations and maps interaction-relevant information to the corresponding widget and workspace representations without modifying the original implementation.
This integration exposes a common interface for reading the current state, invoking actions, issuing perception queries, and accessing interaction history, with communication handled through a message bridge.
Once connected, the existing page state serves as the initial analytical context, while subsequent interactions and observations are mediated through the same widget interface used by natively integrated VA systems.}

\paragraph{\revise{Composing New VA Systems}}
\revise{Users can also compose widgets to build a new VA system directly.
Specifically, they describe the workspace in a specification listing each widget's provider, native view specification, and any links among coordinated views.
The kit then instantiates the corresponding widgets via provider adapters and connects them into a shared workspace.
Similar to Vega and ECharts, we use a declarative workspace specification, which is well suited for assembly.}

To better facilitate human--agent collaborative analysis, we also develop an integrated analysis environment.
Users first upload datasets and visualization specifications in the \textbf{Data Panel} (\Cref{fig:system_overview}A). 
The result is then rendered in the \textbf{Visualization Workspace} (\Cref{fig:system_overview}C), where both human and agent can freely interact and explore.
The \textbf{Analysis Panel} (\Cref{fig:system_overview}B) acts as the conversational and procedural control surface of the system.
\revise{Users switch control between the human and the agent}: humans can annotate insights directly, while the agent plans and acts step by step, with \revise{each iteration shown as an \textit{Observe–Plan–Act–Verify–Reason} card} followed by grounded suggestions for the next step.
The \textbf{Interaction Trace Panel} (\Cref{fig:system_overview}D) visualizes the analysis as a branch-aware provenance graph, where each node is a widget state labeled by its producing action, with human- and agent-initiated nodes visually distinguished.
\revise{Nodes link to their corresponding iteration card and workspace state, so selecting one synchronizes all panels and lets users jump to that point in the history, with additional controls for navigating longer traces.}\looseness=-1

\section{\bench} %  A Benchmark for Agentic Interaction in Visual Analytics
\label{sec:bench}

%\begin{figure*}[t]
%  \centering
%  \includegraphics[width=\textwidth]{figs/visagentbench_construction_draft_compressed.pdf}
%  \caption{\textbf{The construction pipeline of \bench.}
%  We source and construct data-backed workspaces, instantiate them through the widget abstraction, formulate tasks with a target--path taxonomy and an analytical workflow library, and package each task with answer, tool, and state annotations. The benchmark covers single-widget workspaces and coordinated multi-widget workspaces containing two to six widgets.}
%  \label{fig:bench_construction}
%\end{figure*}
Built on this widget abstraction, we construct \bench, a benchmark for evaluating whether agents can \revise{solve VA tasks in both single- and multi-widget workspaces.}
Each instance consists of a workspace, a task prompt, and multiple layers of reference annotations, including a target answer, a reference tool trajectory, and \revise{the expected final workspace state.}
These annotations enable us to evaluate not only the correctness of an agent's conclusion, but also the process by which it reaches that conclusion.
We next describe the construction of the widget pool and task suite, followed by the evaluation protocol.
% Built on our widget abstraction, we construct \bench to systematically evaluate whether an agent can solve VA tasks by operating widgets. 
% \revise{
% The benchmark supports evaluation on both single or multi widgets task.} Each benchmark instance pairs with \revise{one widget or a coordinated workspace containing two to six widgets,} a task prompt and several layers of reference annotations: a target answer
% \revise{the analytical tool trajectory}, and the final \revise{state of the workspace}.
% These layers enable evaluation along complementary dimensions, which allow us to diagnose not only \emph{what} the agent concludes but also \emph{how} it reaches that conclusion.
% We next describe how we construct the widget pool, derive tasks, and define the evaluation protocol.

\subsection{\revise{Workspace Construction}}
\label{sec:bench_widget_construction}

% This section illustrates the process from sourcing a dataset, materializing it into a visualization, and then instantiating it into a widget for analysis. 

\paragraph{\revise{Dataset Sourcing and Selection}}
We construct the benchmark from \revise{a combination of real-world and synthetic datasets.}
For real-world data, we source candidate datasets from Kaggle and retain those that are compatible with at least one supported widget type and provide sufficient analytical structure to support meaningful interaction, excluding datasets whose typical questions can be answered directly from a single static view.
This process yields \revise{96 real-world datasets} \revise{spanning business, public services, education, and other applied domains.}
\revise{To complement these datasets, we additionally construct 304 synthetic datasets with realistic application contexts.
The synthetic datasets allow us to broaden the coverage of data characteristics and analytical relationships while providing controlled and verifiable task settings that are difficult to obtain consistently from independently collected real-world data.}

\paragraph{\revise{Workspace Instantiation}}
%\todo{How to determine the visualization spec for each dataset}
\revise{
For each dataset, we generate the visualization specification using data-type-aware field-mapping rules.
The converter identifies numerical, categorical, and temporal fields and assigns them to the semantic roles required by each widget.
When a benchmark instance requires specific analytical fields, these mappings are provided explicitly instead of using the default
selection.}
%for example, scatterplots preferentially use two numerical fields, while line charts use a temporal or ordered field together with a numerical measure.

\revise{For a single-widget workspace, the resulting visualization specification is rendered and bound as one widget to form one instance.}
\revise{For a multi-widget workspace, we generate the visualization specifications for two to six widgets and instantiate them together as one coordinated workspace. The workspace records the widgets and their coordination relations, and serves as the visual analysis environment of that multi-widget benchmark instance. Additional construction details, including the single-widget field-to-encoding conversion and the multi-widget workflow instantiation procedure, are provided in Supplementary Sec.~S2.1, and the shared benchmark-instance schema is documented in Supplementary Sec.~S2.2.}

\subsection{Task Construction}
\label{sec:bench_taxonomy}

\revise{Given a workspace, we construct analytical tasks that require agents to interact with its visualizations to derive an answer.
To capture different levels of task specification and analytical autonomy, we characterize tasks along the \emph{target} and \emph{path} dimensions following the knowledge-gap formulation of Ceneda~\etal~\cite{ceneda2017characterizingGuidance}, which is also consistent with visualization task typologies that distinguish analytical goals from the means used to achieve them~\cite{Brehmer2013Typology}.
The \emph{target} dimension indicates whether the expected form of the analytical finding is specified, whereas the \emph{path} dimension indicates whether the relevant evidence and major analytical steps are provided.
Their combinations yield four task types, summarized in \Cref{tab:task-taxonomy}.}
We use this taxonomy to guide task construction across workspaces, ensuring systematic coverage of tasks ranging from fully specified analyses to open-ended exploration.
\revise{Specifically, for each workspace, the authors draft seed tasks covering the four task types.}
This design allows the benchmark to assess not only whether an agent can execute prescribed analytical procedures, but also how well it can determine appropriate analytical paths and identify relevant findings when less guidance is provided.

% \subsection{Task Construction}
% \label{sec:bench_taxonomy}
% % Every task instance in VisAgentBench is paired with a unique widget or workspace.
% \revise{Given a workspace, we next construct analytical tasks that require agents to interact with its visualizations to derive an answer.
% To capture tasks with different levels of specification and analytical autonomy, we first define a task taxonomy and use it to guide task generation.}

% \subsubsection{Task Taxonomy}
% we characterize them along \revise{the \emph{target} and \emph{path} dimensions} \revise{following the knowledge-gap characterization proposed by Ceneda et al.~\cite{ceneda2017characterizingGuidance}, which is also consistent with the visualization task typology that distinguish an analytical goal from the methods used to reach it~\cite{Brehmer2013Typology}.
% The \emph{target} dimension indicates whether the expected form of the analytical finding is specified, whereas the \emph{path} dimension indicates whether the relevant evidence and major analytical steps are provided.
% Together, these dimensions yield four task types summarized in Table~\ref{tab:task-taxonomy}.
% }

% \subsubsection{Task Generation}
% To ensure the quality of VA tasks, we construct benchmark tasks through human annotation.
% For each workspace, the authors first draft seed tasks covering the four Target--Path categories

\begin{table*}[t] \centering \caption{Task taxonomy based on target and path specification.} \label{tab:task-taxonomy} \small 
\begin{tabularx}{\textwidth}{llllX} \toprule \textbf{Type} & \textbf{Target} & \textbf{Path} & \textbf{Characteristic} & \textbf{Example} \\ \midrule TK--PK & Known & Known & Specified analysis & Select the largest and smallest wind-farm regions, compare their average vibration values and proportions of high-vibration turbines, and identify the region that performs worse on both indicators. \\\addlinespace[4pt] TK--PU & Known & Unknown & Goal-directed analysis & Between the largest and smallest wind-farm regions, which requires a more urgent maintenance review? Support the answer with evidence. \\\addlinespace[4pt] TU--PK & Unknown & Known & Guided discovery & Inspect region size, vibration distributions, and anomalous turbine records in sequence, and report any noteworthy pattern discovered. \\\addlinespace[4pt] TU--PU & Unknown & Unknown & Open-ended exploration & Explore the wind-turbine dashboard and identify a phenomenon that warrants further investigation. \\ \bottomrule \end{tabularx} \end{table*}

\subsection{Annotation Construction}

\paragraph{Human Annotation}
\label{sec:human}
We recruited eight annotators from a local university, all with computer-science backgrounds and prior experience using visualization for data analysis, to solve the designed tasks through an annotation interface that exposes the same widget operations used during evaluation.
The system records the final answer, interaction trace, final widget or workspace state, and annotator-provided reasoning and key insights.
We convert these records into structured reference annotations for Answer, Reference Trace Similarity, and State evaluation.
This process yields \revise{516 single-widget instances and 608 multi-widget instances.}
Detailed interface implementation and annotation validation procedures are provided in Supplementary Sec.~S2.2.
%\vica{add location}

%The comprehensive benchmark instances schema is stated in Section~\ref{sec:unified_instance_representation} 
% This process yields 516 human-annotated single-widget instances
% and 608 human-annotated multi-widget instances.

\paragraph{Unified Instance Representation}
\label{sec:unified_instance_representation}
All task sources are converted into a shared instance representation.
Each instance contains the analytical query and workspace specification, together with three evaluation-facing annotations: the expected answer or reference insights, \revise{a successful reference interaction trace with accepted alternatives}, and task-relevant final-state checks.
\revise{The instance additionally records its Target--Path type, workspace scope, and provenance.}
This unified representation allows human-authored instances to be evaluated with the same protocol.

\subsection{Evaluation Protocol}
\label{sec:bench_protocol}
% Original: To support systematic development and evaluation of agentic VA behavior, we introduce an evaluation protocol along three dimensions: answer score, reference trace similarity, and state scores.
% Original: A final answer alone cannot reveal whether an agent reached a conclusion through a valid interaction trajectory or happened to produce a plausible response after an unsuccessful interaction, so we introduce a reference trace similarity to evaluate its interaction ability.
% Original: However, matching only one human trajectory exactly would penalize agents that reach the same useful state through another valid tool trajectory, so we introduce final state evaluation: if an agent implements another reasonable tool trajectory but reaches a state that is adequate to answer the query successfully, it receives a high score in the state dimension.
\revise{A single end-task accuracy score cannot fully reflect an agent's performance, since it conflates a correct answer with the validity of the interaction process that produced it.
Because each task instance already packages an expected answer, a reference tool trace, and task-relevant state conditions, we evaluate along three matching dimensions: Answer, Reference Trace Similarity, and State.
Reference Trace Similarity checks whether the agent's interaction trace is consistent with a valid evidence-gathering process, so a plausible answer produced without a supporting trace is not mistaken for genuine interaction competence.
However, scoring the trace alone against one human-annotated reference would penalize an agent that reaches an equally useful state through a different, still-valid trace, so State credits any trace whose final workspace state is adequate to answer the query, independent of how that state was reached.}

% (\Cref{tab:instance_schema})

\paragraph{Answer Score}
% Original: For clear tasks with deterministic ground-truth answers, we evaluate the final response with answer-type-specific matching.
The Answer Score evaluates the final response \revise{against each task's expected answer using an evaluation procedure suited to the task type.}
For tasks with deterministic ground-truth answers, we apply answer-format-specific matching.
% \revise{The evaluator supports five answer types: numeric, boolean, categorical, interval, and open-ended insight.}
Specifically, numeric answers are checked with a tolerance threshold, boolean answers with normalized string matching~\cite{masry2025chartqaprodiversechallengingbenchmark}, and categorical answers with exact matching plus an LLM judge for a semantic-equivalence check~\cite{wang2024charxivchartinggapsrealistic,lu2024mathvistaevaluatingmathematicalreasoning}.
% \revise{For interval, date intervals require exact agreement between both bounds, while numeric intervals allow the tolerance specified by the instance.}
\revise{For Target-Unknown (TU) tasks, we score against a set of reference insights.
The evaluator computes claim precision $P$, claim recall $R$, and uses a rubric to score groundedness $G$, which assesses how faithfully the response preserves available analytical evidence.
The detailed rubric is presented in Supplementary Sec.~S2.3.1.
%\vica{add location}
The score is calculated as}
\begin{equation}
    S_{\mathrm{answer}}
    =
    \frac{2PR}{P+R}\cdot G.
\end{equation}
\revise{We use \textbf{Claude Sonnet 4.6~\cite{anthropic2026claudesonnet46}} as the judge model and compare its scores with two independent human raters on open-ended tasks~\cite{masry2022chartqa,min2023factscore,liu2023geval,zheng2023llmJudge}.
The judge reaches ICC(2,1) values of \(0.76\) and \(0.73\) with the two raters, showing good agreement with human ratings. The details are presented in \Cref{sec:agreement}.}

%The detailed results are presented in the supplement material.
%\vica{add claim the results. details are in supp}
%\vica{How G is calculated? What is the scope? Is it reliable? and what is the agreement here.}
%by a rubric, this would be presented in the supplement materials.

\paragraph{\revise{Reference Trace Similarity}}
This dimension \revise{evaluates whether the agent completes the required interaction steps in an order consistent with the reference trace.
% Only successful tool calls are considered.
Each required reference step specifies an operation, a target widget, the parameters, and additionally provides accepted alternative operations.
}
\revise{Because an agent may issue exploratory calls that are not part of the reference trace, comparing the two sequences position by position would unfairly penalize otherwise correct behavior.}
We therefore align the agent's successful calls with the reference steps \revise{in the spirit of longest-common-subsequence matching:} each reference step may be matched to at most one agent call, matched pairs must preserve the original order of the reference steps, \revise{and any extra agent calls between two matches are simply skipped rather than penalized.}
\revise{Among all such order-preserving alignments, we take the one that maximizes the total similarity between matched pairs, and report the trace score as this maximum total similarity divided by the number of required reference steps.}
Optional reference steps and unmatched agent calls are reported separately for diagnostic purposes but do not affect the score.

\paragraph{State Score}
This dimension assesses whether the executed interactions produce a final state that appropriately supports the task, rather than rewarding an interaction trace that looks plausible but leaves the visualization in an ineffective or incorrect state.
\revise{State Score and Reference Trace Similarity are therefore complementary rather than redundant: Reference Trace Similarity checks whether the agent's process resembles a valid evidence-gathering procedure, while State Score checks whether the agent actually ended up in a state that supports the task, regardless of which procedure it followed to get there.}
% For example, consider a scatter-plot task where the agent must zoom into a dense region before answering. The initial state leaves both axes at the full domain, whereas the target state should constrain the visible ranges to the region of interest.
% Original: State checks cover widget selections, data transforms, and view properties. For each applicable check $q_k$, the evaluator reads the corresponding observed value $\hat{q}_k$ and applies a corresponding matching rule. Structured states support subset-style matching, while numeric or range-valued properties may use annotated tolerances. Each check receives a binary score.
Annotators specify the conditions that the final state should satisfy to solve the task, and the evaluator checks whether the final state satisfies these conditions and reports the fraction of the conditions that are satisfied.

\section{Evaluation}
\label{sec:evaluation}
% Original: To validate the widget abstraction framework, we evaluate whether current VLM agents can complete VA tasks under different levels of context, examining how different settings affect VA task performance.
We evaluate the effectiveness of the framework on \bench by comparing \revise{four execution settings that isolate tool access, widget abstraction, and workflow guidance.}
The comparison is also \revise{complemented by a human reference study, two usage scenarios, and an analysis of failure modes.}

% The single-widget results are shown in
% \cref{tab:single_setting_results}.

%     \endgroup
% \end{table}

\subsection{Benchmark Evaluation}
\label{sec:experimental_setup}

\subsubsection{Evaluation Setup}
\paragraph{Baselines}
% Original: We evaluate a set of state-of-the-art VLM agents including proprietary and open-source families, all operating within the same agentic loop introduced in \Cref{sec:agentic_orchestration}.
% Original: Proprietary models include Gemini 3.6 Flash, GPT-5.6-Luna. Open-source models use Qwen3-vl-30b-a3b-instruct.
We evaluate three VLM backbones covering proprietary and open-source families \revise{(Gemini 3.6 Flash~\cite{google2026gemini36flash}, GPT-5.6-Luna~\cite{openai2026gpt56luna}, and Qwen3-VL-30B-A3B-Instruct~\cite{qwen2025qwen3vl30ba3b})}.
\revise{The Claude family is not included because it is used as the judge model for open-ended tasks. }
%(\Cref{xxx}).
% Original: We also include a human baseline for interpreting agent performance.
% Original: To provide a human reference for interpreting agent performance, we sample 100 benchmark tasks as a validation subset, including 50 single-widget and 50 multi-widget tasks, with balanced coverage of the four Target-Path task types. We recruit 2 participants to complete the tasks through our benchmark annotation system. Annotation-specific information, including reference answers and reference traces, is hidden from the participants. The system records each participant's final answer, interaction log, and final application state.
\revise{We also include a human baseline on a 10\% subset with balanced coverage of the different task types}.
\revise{Two participants complete the tasks through the benchmark annotation system, and the system records each participant's final answer, interaction log, and final application state for evaluation.}

\paragraph{Agentic Execution Settings}
% Original: We compare 4 settings that isolate the effect of our proposed widget framework.
% Original: In No Tools setting (S0), the model receives the task query and the initial view context but cannot invoke tools.
% Original: In Direct Tools setting (S1), the model has access to a set of widget tools through multi-turn interaction.
% Original: In Widget Abstraction setting (S2), the model has access to the same interactive tools, but these tools are organized through the widget-level state-action-perception abstraction described in Sec. framework_widget.
% Original: In Widget Abstraction + Workflow setting (S3), we additionally provide workflow for the model to organize VA analysis.
We compare four settings that isolate the effect of the \revise{widget abstraction framework:}
\begin{itemize}
    \item \revise{$S_0$ (\textbf{No Tools}):} the model receives the task query and the initial view context but cannot invoke tools.
    \revise{
    \item $S_1$ (\textbf{Direct Tools}): the model has access to a set of widget tools through multi-turn interaction.
    \item $S_2$ (\textbf{Widget Abstraction}): the model has access to the same interactive tools, but these tools are organized through the widget-level state--action--perception abstraction described in \Cref{sec:framework_widget}.
    \item $S_3$ (\textbf{Widget Abstraction + Workflow}): we additionally provide workflows to structure the VA analysis process.}
\end{itemize}

\paragraph{\revise{Environment Setup}}
\revise{All experiments were executed using the benchmark's standard evaluation runner.
Each instance was launched as an independent session, preventing interaction state from carrying over across runs.
Agents were given a maximum budget of 16 turns for each task.  
% Model inference was performed remotely through the API; the local machine was used only to schedule runs, render views, and store execution traces and evaluation outputs. We used the same benchmark instances, evaluation protocol, and per-level prompt templates for all models.
}
%The benchmark runner, instance specifications, and evaluation scripts are provided in the supplementary material.

%TODO update supplement

%\paragraph{Human validation of LLM-based Answer scoring.}
%\revise{For open-ended and some categorical answers that cannot be evaluated by exact matching, we use Claude Sonnet 4.6 as an LLM judge. To validate agreement between automatic scores and human judgment, we independently sample 200 model responses scored by the judge from the validation subset, covering the three models, four execution settings, single- and multi-widget tasks, and the four Target-Path task types, with emphasis on open-ended and categorical responses.}

%\revise{Two human raters independently evaluate all sampled responses using the same answer evaluation criteria as the judge. The raters are blinded to model identity, execution setting, and the judge's scores. We measure
%inter-rater agreement using ICC and compare Claude scores with the mean human ratings using Spearman's correlation and mean absolute error (MAE). The two raters achieve an ICC of \(\); Claude scores correlate with mean human ratings at \(\rho=\), with an MAE of \(\). }

\begin{table*}[t]
    \caption{
    Experimental results for single- and multi-widget tasks.
    % All values are percentages.
    N/A indicates that Trace and State are not applicable under $S_0$.
    % Human results are reported as overall reference scores.
    }
    \label{tab:setting_results}
    \label{tab:single_setting_results}
    \label{tab:multi_setting_results}

    \centering
    \begingroup

    %\normalfont
    %\rmfamily

    % Narrower horizontally and slightly more spacious vertically.
    \setlength{\tabcolsep}{7.0pt}
    \renewcommand{\arraystretch}{1.0}

    \setlength{\aboverulesep}{0.25ex}
    \setlength{\belowrulesep}{0.25ex}
    \setlength{\cmidrulesep}{0.20ex}

    \begin{tabular}{
        c
        !{\color{singleheatblue!32}\vrule width 0.35pt}
        *{3}{c}
        !{\color{singleheatblue!32}\vrule width 0.35pt}
        *{3}{c}
        !{\color{singleheatblue!32}\vrule width 0.35pt}
        *{3}{c}
        !{\color{singleheatblue!32}\vrule width 0.35pt}
        *{3}{c}
        !{\color{singleheatblue!22}\vrule width 0.30pt}
        l
    }

        % ====================================================
        % Header
        % ====================================================

        \toprule

        \multicolumn{1}{c}{} &
        \multicolumn{3}{c}{\textbf{TK-PK}} &
        \multicolumn{3}{c}{\textbf{TK-PU}} &
        \multicolumn{3}{c}{\textbf{TU-PK}} &
        \multicolumn{3}{c}{\textbf{TU-PU}} &
        \multicolumn{1}{c}{}
        \\[-1pt]

        \cmidrule(lr){2-4}
        \cmidrule(lr){5-7}
        \cmidrule(lr){8-10}
        \cmidrule(lr){11-13}

        \textbf{Setting} &
        \textbf{Answer} & \textbf{Trace} & \textbf{State} &
        \textbf{Answer} & \textbf{Trace} & \textbf{State} &
        \textbf{Answer} & \textbf{Trace} & \textbf{State} &
        \textbf{Answer} & \textbf{Trace} & \textbf{State} &
        \textbf{Model (Avg. A)}
        \\

        \midrule

        % ====================================================
        % Single-widget
        % ====================================================

        \multicolumn{14}{l}{\textbf{Single-widget tasks}}
        \\
        \midrule

        % -------------------- S0 --------------------

        \singlestatelabel{S_0}
        & \singleheat{52.5} & \textit{N/A} & \textit{N/A}
        & \singleheat{56.4} & \textit{N/A} & \textit{N/A}
        & \singleheat{35.2} & \textit{N/A} & \textit{N/A}
        & \singleheat{35.9} & \textit{N/A} & \textit{N/A}
        & \textbf{GPT} (45.0)
        \\

        &
        \singleheat{54.2} & \textit{N/A} & \textit{N/A}
        & \singleheat{56.7} & \textit{N/A} & \textit{N/A}
        & \singleheat{37.3} & \textit{N/A} & \textit{N/A}
        & \singleheat{34.5} & \textit{N/A} & \textit{N/A}
        & \textbf{Gemini} (45.7)
        \\

        &
        \singleheat{33.3} & \textit{N/A} & \textit{N/A}
        & \singleheat{31.6} & \textit{N/A} & \textit{N/A}
        & \singleheat{16.2} & \textit{N/A} & \textit{N/A}
        & \singleheat{23.0} & \textit{N/A} & \textit{N/A}
        & \textbf{Qwen} (26.0)
        \\

        \cmidrule(lr){1-13}

        % -------------------- S1 --------------------

        \singlestatelabel{S_1}
        & \singleheat{70.8} & \singleheat{68.1} & \singleheat{67.2}
        & \singleheat{63.7} & \singleheat{49.1} & \singleheat{39.0}
        & \singleheat{64.5} & \singleheat{58.3} & \singleheat{55.5}
        & \singleheat{46.9} & \singleheat{37.8} & \singleheat{34.3}
        & \textbf{GPT} (61.4)
        \\

        &
        \singleheat{59.4} & \singleheat{62.9} & \singleheat{67.2}
        & \singleheat{57.3} & \singleheat{44.7} & \singleheat{39.8}
        & \singleheat{43.0} & \singleheat{55.0} & \singleheat{53.3}
        & \singleheat{39.8} & \singleheat{31.0} & \singleheat{33.3}
        & \textbf{Gemini} (49.9)
        \\

        &
        \singleheat{36.2} & \singleheat{44.9} & \singleheat{48.0}
        & \singleheat{37.9} & \singleheat{32.7} & \singleheat{29.8}
        & \singleheat{21.6} & \singleheat{41.1} & \singleheat{40.3}
        & \singleheat{27.7} & \singleheat{21.1} & \singleheat{18.7}
        & \textbf{Qwen} (30.9)
        \\

        \cmidrule(lr){1-13}

        % -------------------- S2 --------------------

        \singlestatelabel{S_2}
        & \singleheat{78.2} & \singleheat{73.6} & \singleheat{72.4}
        & \singleheat{76.3} & \singleheat{60.5} & \singleheat{50.0}
        & \singleheat{65.0} & \singleheat{66.9} & \singleheat{60.1}
        & \singleheat{51.2} & \singleheat{42.8} & \singleheat{40.5}
        & \textbf{GPT} (67.7)
        \\

        &
        \singleheat{62.9} & \singleheat{66.7} & \singleheat{74.0}
        & \singleheat{67.7} & \singleheat{52.2} & \singleheat{45.6}
        & \singleheat{46.0} & \singleheat{56.4} & \singleheat{56.8}
        & \singleheat{46.1} & \singleheat{33.3} & \singleheat{38.1}
        & \textbf{Gemini} (55.6)
        \\

        &
        \singleheat{37.4} & \singleheat{50.8} & \singleheat{56.1}
        & \singleheat{33.2} & \singleheat{36.0} & \singleheat{32.5}
        & \singleheat{24.7} & \singleheat{45.4} & \singleheat{42.5}
        & \singleheat{25.5} & \singleheat{26.6} & \singleheat{24.5}
        & \textbf{Qwen} (30.2)
        \\

        \cmidrule(lr){1-13}

        % -------------------- S3 --------------------

        \singlestatelabel{S_3}
        & \singleheat{84.3} & \singleheat{80.7} & \singleheat{73.5}
        & \singleheat{80.7} & \singleheat{73.0} & \singleheat{62.4}
        & \singleheat{68.8} & \singleheat{78.0} & \singleheat{69.8}
        & \singleheat{56.6} & \singleheat{66.4} & \singleheat{63.1}
        & \textbf{GPT} (72.6)
        \\

        &
        \singleheat{76.2} & \singleheat{76.0} & \singleheat{73.8}
        & \singleheat{65.3} & \singleheat{66.2} & \singleheat{59.5}
        & \singleheat{48.8} & \singleheat{69.6} & \singleheat{69.5}
        & \singleheat{44.3} & \singleheat{55.7} & \singleheat{53.8}
        & \textbf{Gemini} (58.7)
        \\

        &
        \singleheat{46.3} & \singleheat{67.7} & \singleheat{69.6}
        & \singleheat{43.0} & \singleheat{50.5} & \singleheat{50.8}
        & \singleheat{30.3} & \singleheat{59.6} & \singleheat{57.2}
        & \singleheat{27.5} & \singleheat{44.3} & \singleheat{45.1}
        & \textbf{Qwen} (36.8)
        \\

        \midrule

        \multicolumn{1}{l}{\textbf{}}
        & \singleheat{83.6} & \singleheat{84.8} & \singleheat{71.7}
        & \singleheat{79.4} & \singleheat{76.1} & \singleheat{64.6}
        & \singleheat{69.7} & \singleheat{74.3} & \singleheat{59.3}
        & \singleheat{66.5} & \singleheat{64.4} & \singleheat{52.7}
        & \textbf{Human} (74.8)
        \\

        % ====================================================
        % Multi-widget
        % ====================================================

        \midrule
        \multicolumn{14}{l}{\textbf{Multi-widget tasks}}
        \\
        \midrule

        % -------------------- S0 --------------------

        \singlestatelabel{S_0}
        & \singleheat{67.8} & \textit{N/A} & \textit{N/A}
        & \singleheat{61.8} & \textit{N/A} & \textit{N/A}
        & \singleheat{29.8} & \textit{N/A} & \textit{N/A}
        & \singleheat{28.6} & \textit{N/A} & \textit{N/A}
        & \textbf{GPT} (47.0)
        \\

        &
        \singleheat{57.9} & \textit{N/A} & \textit{N/A}
        & \singleheat{59.9} & \textit{N/A} & \textit{N/A}
        & \singleheat{20.2} & \textit{N/A} & \textit{N/A}
        & \singleheat{18.6} & \textit{N/A} & \textit{N/A}
        & \textbf{Gemini} (39.2)
        \\

        &
        \singleheat{51.3} & \textit{N/A} & \textit{N/A}
        & \singleheat{42.8} & \textit{N/A} & \textit{N/A}
        & \singleheat{13.2} & \textit{N/A} & \textit{N/A}
        & \singleheat{12.5} & \textit{N/A} & \textit{N/A}
        & \textbf{Qwen} (30.0)
        \\

        \cmidrule(lr){1-13}

        % -------------------- S1 --------------------

        \singlestatelabel{S_1}
        & \singleheat{71.7} & \singleheat{55.0} & \singleheat{70.4}
        & \singleheat{77.6} & \singleheat{25.9} & \singleheat{25.7}
        & \singleheat{24.6} & \singleheat{50.4} & \singleheat{67.8}
        & \singleheat{29.8} & \singleheat{4.5} & \singleheat{6.0}
        & \textbf{GPT} (51.0)
        \\

        &
        \singleheat{58.6} & \singleheat{46.2} & \singleheat{53.3}
        & \singleheat{60.5} & \singleheat{15.8} & \singleheat{11.8}
        & \singleheat{13.1} & \singleheat{44.9} & \singleheat{54.7}
        & \singleheat{17.7} & \singleheat{1.4} & \singleheat{1.3}
        & \textbf{Gemini} (37.6)
        \\

        &
        \singleheat{54.6} & \singleheat{38.8} & \singleheat{36.5}
        & \singleheat{46.7} & \singleheat{8.1} & \singleheat{9.9}
        & \singleheat{9.1} & \singleheat{35.3} & \singleheat{36.9}
        & \singleheat{9.2} & \singleheat{2.4} & \singleheat{2.0}
        & \textbf{Qwen} (30.0)
        \\

        \cmidrule(lr){1-13}

        % -------------------- S2 --------------------

        \singlestatelabel{S_2}
        & \singleheat{88.2} & \singleheat{75.3} & \singleheat{94.6}
        & \singleheat{78.3} & \singleheat{35.6} & \singleheat{37.7}
        & \singleheat{33.3} & \singleheat{76.5} & \singleheat{90.4}
        & \singleheat{32.7} & \singleheat{10.3} & \singleheat{13.5}
        & \textbf{GPT} (58.1)
        \\

        &
        \singleheat{59.2} & \singleheat{68.2} & \singleheat{96.0}
        & \singleheat{64.5} & \singleheat{28.5} & \singleheat{26.9}
        & \singleheat{22.4} & \singleheat{71.5} & \singleheat{94.1}
        & \singleheat{25.2} & \singleheat{2.4} & \singleheat{2.8}
        & \textbf{Gemini} (43.0)
        \\

        &
        \singleheat{61.8} & \singleheat{55.1} & \singleheat{75.9}
        & \singleheat{52.0} & \singleheat{19.4} & \singleheat{24.3}
        & \singleheat{11.9} & \singleheat{54.2} & \singleheat{75.8}
        & \singleheat{13.8} & \singleheat{13.2} & \singleheat{20.8}
        & \textbf{Qwen} (35.0)
        \\

        \cmidrule(lr){1-13}

        % -------------------- S3 --------------------

        \singlestatelabel{S_3}
        & \singleheat{88.8} & \singleheat{67.6} & \singleheat{90.1}
        & \singleheat{81.6} & \singleheat{39.9} & \singleheat{40.7}
        & \singleheat{36.9} & \singleheat{71.6} & \singleheat{88.2}
        & \singleheat{35.3} & \singleheat{34.7} & \singleheat{36.1}
        & \textbf{GPT} (60.7)
        \\

        &
        \singleheat{63.8} & \singleheat{68.0} & \singleheat{93.5}
        & \singleheat{66.4} & \singleheat{32.6} & \singleheat{36.1}
        & \singleheat{23.2} & \singleheat{69.5} & \singleheat{89.3}
        & \singleheat{23.2} & \singleheat{26.5} & \singleheat{29.7}
        & \textbf{Gemini} (44.2)
        \\

        &
        \singleheat{63.8} & \singleheat{53.1} & \singleheat{76.1}
        & \singleheat{57.2} & \singleheat{20.1} & \singleheat{23.6}
        & \singleheat{12.8} & \singleheat{52.1} & \singleheat{70.4}
        & \singleheat{12.8} & \singleheat{18.1} & \singleheat{23.2}
        & \textbf{Qwen} (36.8)
        \\

        \midrule

        \multicolumn{1}{l}{\textbf{}}
        & \singleheat{76.9} & \singleheat{56.2} & \singleheat{70.8}
        & \singleheat{91.7} & \singleheat{40.0} & \singleheat{56.3}
        & \singleheat{26.3} & \singleheat{39.2} & \singleheat{56.1}
        & \singleheat{33.9} & \singleheat{44.9} & \singleheat{53.8}
        & \textbf{Human} (57.1)
        \\

        \bottomrule

    \end{tabular}

    \endgroup
\end{table*}

\subsubsection{Result Analysis}
\label{sec:result_analysis}
\label{sec:single_results}
\label{sec:multi_results}

% Original: Cref{tab:setting_results} reports Answer, Trace, and State scores across models, settings, and Target--Path types.
\revise{\Cref{tab:setting_results} reports Answer, Trace, and State scores across models, settings, and Target--Path types.}
\revise{The four settings isolate tool access, widget organization, and workflow guidance, so we read $S_1$--$S_2$ and $S_2$--$S_3$ as interface effects rather than as changes in what the agent can execute.}

\paragraph{\revise{Single-Widget Tasks}}
\revise{All three models obtain their strongest overall scores under $S_3$.}
\revise{Moving from No Tools to Direct Tools improves Answer, as expected: many tasks require evidence that is not visible in the initial view, so the agent has to interact before it can form a correct conclusion.}
\revise{The more diagnostic comparison is $S_1$--$S_2$, where the executable tools are held fixed.}
\revise{Widget Abstraction raises Answer for GPT and Gemini and slightly lowers it for Qwen, while Qwen's Trace and State still rise.}
\revise{An explanation is that the widget abstraction makes the current state, applicable actions, and perception queries easier to select, so Qwen can ground interactions more reliably without yet mapping the acquired evidence onto the required answer.}
\revise{This reading is consistent with the evidence-to-answer errors reported later (\Cref{sec:failure_modes}), which remain more frequent for Qwen than for GPT even under $S_3$.}
\revise{Workflow then improves Trace and State more than Answer.}
\revise{The pattern suggests that reusable procedures mainly help the agent choose and sequence operations when the path is underspecified, whereas synthesizing a correct final answer still depends more on the model capability.}
\revise{Widget abstraction and workflow are therefore complementary: the former organizes what the agent can do in the current view, and the latter organizes what it should do next.}

\paragraph{\revise{Multi-Widget Tasks}}
\revise{Multi-widget tasks do not follow the same progression.}
\revise{Direct Tools add little to Answer, and Gemini's overall Answer even decreases slightly relative to No Tools.}
\revise{A likely reason is that a coordinated workspace exposes many more operations than a single widget, so a flat tool list increases selection difficulty without making cross-view state changes attributable to the originating action.}
\revise{The main step is therefore $S_1$--$S_2$: averaged across models, Answer rises by $5.8$ points, while Trace and State rise by $15.1$ and $23.1$ points.}
\revise{The much larger interaction gains indicate that agents already had the needed operations, but lacked an inspectable account of widget state and of how an action in one view affects another.}
\revise{Workflow adds a smaller further gain, suggesting that once coordination is inspectable, additional path templates help less than they do on single-widget tasks.}
\revise{Where Trace and State already rise sharply, the remaining Answer gap is more consistent with insight selection than with missing step order.}
\revise{Across both scopes, target-unknown tasks remain harder on Answer, because the agent must decide which finding is worth reporting rather than match a specified target.}
\revise{Path-unknown tasks remain harder on Trace and State, because the agent must discover a useful interaction sequence rather than follow a stated procedure.}
\revise{Both difficulties concentrate on multi-widget TU-PU, where the agent has to choose what to look for and how to look across views at the same time.}

% Original: For the Multi-Widget evaluation, we assessed result stability using a fixed subset of 10% instances, corresponding to 10% of the full 608-instance Multi-Widget benchmark. GPT-5.6 Luna, Gemini-3.6 Flash, and Qwen3-VL-30B-A3B-Instruct were evaluated under four Planner settings across three aligned result sets, yielding 2,196 model--Planner--instance evaluations in total. The results exhibit moderate and generally limited variation (Answer: median SD =2.7%, mu_SD=2.7%; Trace: median SD =1.8%, mu_SD=1.5%; State: median SD =1.9%, mu_SD=1.8%). For the 11 instances added to reach approximately 10% coverage, one result set reuses the corresponding formal benchmark results, while the other two were obtained from independent executions. Complete configuration-level results and 95% paired-instance bootstrap confidence intervals are reported in the supplemental material.

% For multi-widget tasks, we assess stability on a 10\% subset of the multi-widget tasks.
% The three models are evaluated under the four settings across three aligned result sets.
% The results exhibit limited variation (Answer: median SD \(=2.7\%\), \(\mu_{\mathrm{SD}}=2.7\%\); Trace: median SD \(=2.0\%\), \(\mu_{\mathrm{SD}}=2.0\%\); State: median SD \(=2.2\%\), \(\mu_{\mathrm{SD}}=2.4\%\)).
% Complete configuration-level means and sample standard deviations across the three repetitions are reported in the supplemental material.

\paragraph{\revise{Relative to the Human Baseline}}
\revise{On single-widget tasks, agents remain below the human Answer reference, so interface support improves operability without closing the gap in conclusion quality.}
\revise{On multi-widget tasks, GPT under $S_3$ reaches an Answer score of 60.7\%, slightly above the human reference of 57.1\%.
This observation is consistent with the broader pattern reported above:
multi-widget tasks show the largest gains from $S_1$ to $S_2$, indicating that structured widget support is particularly useful when analysis requires coordination across multiple views.}
\revise{
This difference in evidence-acquisition strategy is also reflected in the human evaluation scores.
The human Trace score is lower than the State score because participants often manipulate the visualization and interpret the view directly, rather than invoking the perception queries recorded in the reference trace.}
\revise{
In contrast, perception queries account for $41.0\%$, $29.2\%$, and $33.9\%$ of successful calls under $S_3$ for GPT, Gemini, and Qwen, respectively.
The more explicit use of structured perception may help agents maintain and integrate evidence across coordinated views, particularly in multi-widget analysis.
}
%\revise{The comparison does not imply that GPT is a stronger analyst than the human participants.}
%\revise{The human Trace score is lower than State, and humans are more often to manipulate the visualization and read the updated view.}
%issue explicit perception queries in only $9.3\%$ of operations:
%\revise{State can still credit that strategy, but Trace penalizes it when the reference records perception calls.}
%\revise{Under $S_3$, perception accounts for $41.0\%$, $29.2\%$, and $33.9\%$ of successful calls for GPT, Gemini, and Qwen.}
%\revise{Agents therefore retrieve evidence through structured perception queries rather than by inspecting the updated view as humans do, which can raise Trace, and sometimes Answer, without implying a more human-like analysis strategy.}

\paragraph{\revise{Agreement with Human Ratings}}
\label{sec:agreement}
\revise{To assess the reliability of the LLM-based evaluator for open-ended tasks, we compare Claude Sonnet 4.6 scores with two independent human raters on a sample of 200 responses. Claude achieved ICC(2,1) values of \(0.76\) and \(0.73\) and mean absolute errors (MAEs) of \(0.12\) and \(0.14\), respectively.
These results support the reliability of Claude-based scoring for open-ended tasks. More details are reported in Supplementary Sec.~S2.3.4.}
% Human-human agreement is higher, with  ICC(2,1)\(=0.87\) and an MAE of \(0.10\) . 
% and Spearman correlations of \(0.74\) and \(0.71\)

\paragraph{\revise{Stability Analysis}}
\label{sec:stability}
\revise{To assess result stability, we repeated each model under all four settings three times on fixed 10\% subsets of both the single- and multi-widget benchmarks.
Across the repeated runs, the results exhibited limited variation, with an overall average standard deviation of approximately \(2.4\) points.
%\vica{check the result}
Complete repeated-run results, including the mean and sample standard deviation for each model, execution setting, and workspace scope, are
reported in Supplementary Sec.~S3.2.}
%\vica{add location}}

%The other detailed results are presented in the supplementary material.

\subsection{Usage Scenarios}
\label{sec:user_scenario}

To complement the quantitative evaluation, we present two \revise{usage scenarios} that \revise{illustrate how the widget abstraction} supports practical visual analysis in the two integration settings described in \Cref{sec:framework_implementation}.

\subsubsection{Scenario 1: Operational Risk Analysis}
\label{sec:case_autonomous}

\revise{This scenario illustrates how users and agents conduct collaborative analysis in our integrated analysis environment.}

\paragraph{Open-Ended Operational Triage}
\revise{Daniel worked through a service-reliability triage task in which he needed to determine which operational regions needed deeper investigation.}
The bubble scatterplot encoded latency on the x-axis, error rate on the y-axis, traffic volume by bubble size, and operational region by color.
\revise{The initial overview contained both a dense central band and several distant outlier groups.
Although the chart suggested that operational risk increased toward the upper-right region, the outliers stretched the visual scale and made it difficult to determine whether the dense center contained distinct regional risk profiles.}
\revise{Rather than manually testing a sequence of filters and zoom regions, Daniel switched the Analysis Panel to Agent mode and issued an open-ended request:} ``Please explore this scatterplot autonomously and summarize risk patterns.'' \revise{The agent subsequently expressed its exploration through semantic widget actions such as filtering and zooming (R1).}

\revise{The iteration cards exposed the agent's current action and next analytical sub-goal, allowing Daniel to understand how the exploration was being organized rather than receiving only its final result (\Cref{fig:system_overview} 1.1--1.2).}

\paragraph{View Refinement and Verification}

The agent first applied \revise{\texttt{scatter.filterCategorical}} to remove \textit{Legacy-Core} and \revise{\textit{Factory-IoT}}, \revise{the two most isolated outlier regions in the initial view (\Cref{fig:system_overview} 2.1).}
\revise{This operation reduced the visual influence of the extreme points and made the dense operating region easier to inspect, while retaining \textit{Global-Backbone} as a potentially relevant secondary risk group.}
Based on the filtered state, the agent then applied \revise{\texttt{scatter.zoomDomain}} to \revise{focus on the dense center of the chart (\Cref{fig:system_overview} 2.2).}
The Verification panel confirmed that the filtering had been executed successfully and that the resulting state provided a new and deeper perspective for the analysis (\Cref{fig:system_overview} 2.3).

\revise{
Daniel could inspect the updated views and how they were produced.
The Interaction Trace recorded the transition from the initial view to the filtered and zoomed states (\Cref{fig:system_overview} D1).
Each trace node corresponded to an iteration card in the Analysis Panel and to the resulting state in the Visualization Workspace (\Cref{fig:system_overview} D3).
Daniel could therefore revisit a step and inspect its plan, executed operation, and visual outcome together, rather than reconstructing the analysis from a disconnected sequence of interface changes (R2). }

\paragraph{Findings and Follow-Up}
In the zoomed view, three regional risk tiers became distinguishable: \revise{\textit{Central-Cloud} formed a lower-risk tier, \textit{Edge-Store} a moderate-risk tier, and \textit{Healthcare-Edge} a higher-risk tier.}
The refined view also showed a clear gradient in which higher latency and error rates were associated with increased risk, with larger-traffic operations concentrated toward the higher-risk portion of the chart.
The agent summarized these findings in the final response, \revise{grounding its claims in the state produced through the preceding filtering and zooming operations (\Cref{fig:system_overview} 2.4).
For Daniel, this transformed the original cluttered overview into a more concrete basis for deciding which service regions and high-traffic operations merited further examination.}

The analysis did not end with a static summary.
Based on the current findings, the system proposed four follow-up analyses: comparing the three identified tiers in terms of latency and traffic, examining \textit{Global-Backbone} as a secondary risk group, identifying high-traffic outliers, and comparing the previously filtered \textit{Legacy-Core} and \textit{Factory-IoT} regions (\Cref{fig:system_overview} 3.1).
Because these suggestions were grounded in the \revise{current workspace state, Daniel could continue directly from the refined view without recreating the filters, zoom region, or analytical context.}
\revise{The system thus helped him move from an ambiguous overview to an inspectable set of findings and a reusable operational triage plan.}

\begin{figure*}[t]
    \centering
    \includegraphics[width=\textwidth]{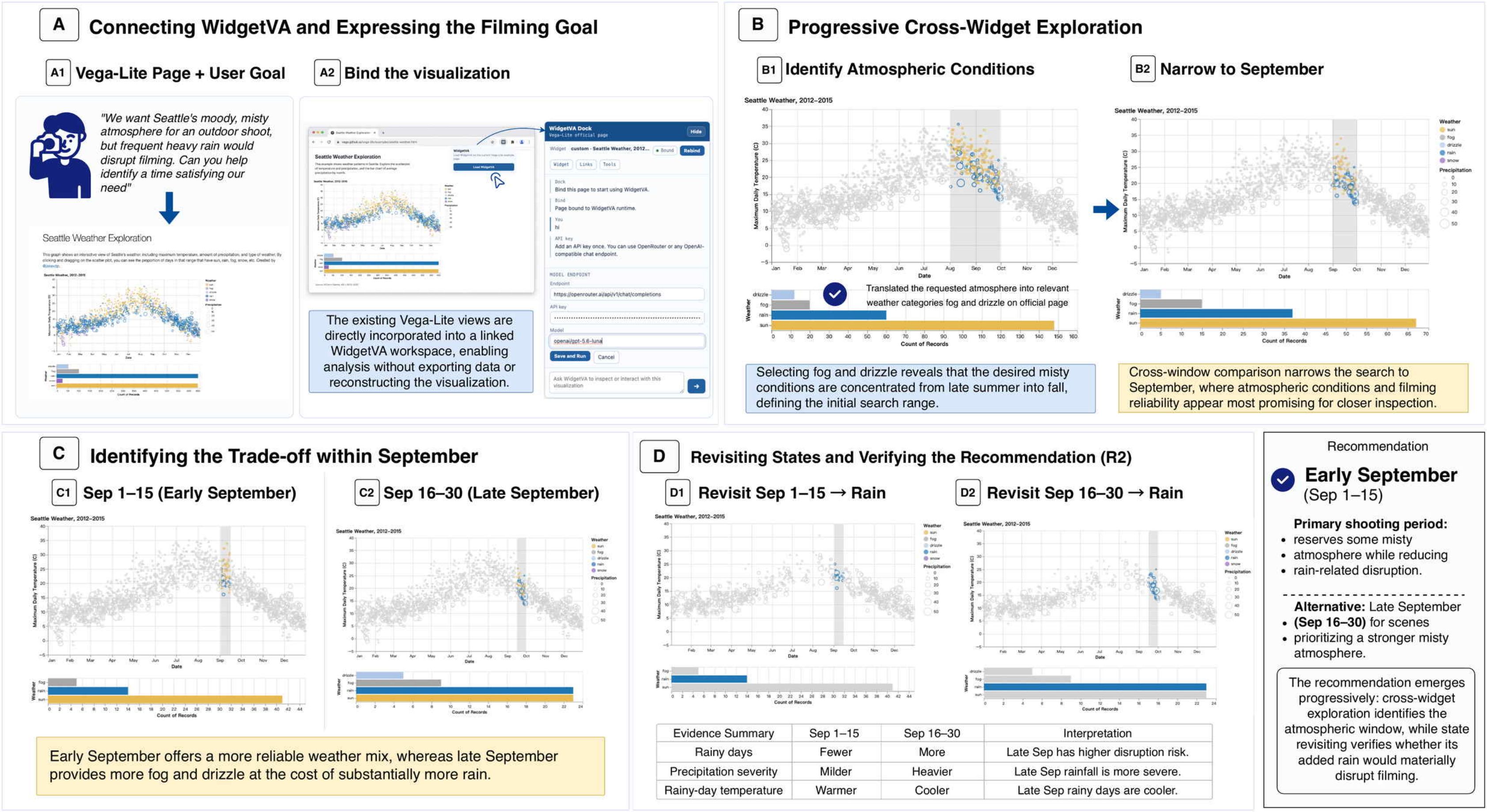}
    \caption{
    \revise{
    Scenario 2: finding an atmospheric filming window on an existing Vega-Lite page.
    (A) WidgetVA is connected directly to the existing visualization and receives the user's filming goal (R3).
    (B) Cross-widget filtering and temporal brushing identify the relevant atmospheric conditions and progressively narrow the candidate period to September (R1).
    (C) Comparing early and late September reveals an atmosphere--reliability trade-off: late September provides more fog and drizzle but also substantially more rain.
    (D) The agent revisits the saved September states and drills down into rainy-day records (R2), verifying the higher disruption risk in late September and supporting Sept.~1--15 as the recommended filming window.
    }}
    \label{fig:official_page_case}
\end{figure*}

\subsubsection{\revise{Scenario 2: Finding an Atmospheric Filming Time}}
\revise{ \label{sec:case_official_page}}

\revise{This scenario illustrates how the same widget abstraction can be applied to an existing VA page and support state-preserving cross-widget analysis.}
\revise{\paragraph{Connecting WidgetVA and Expressing the Filming Goal}
Ethan was a photographer preparing an outdoor shoot in Seattle.
She wanted to show the city's characteristic moody and misty atmosphere, but frequent or heavy rain would interrupt filming.
Therefore, she needed to find a suitable filming period that preserved the desired atmosphere while avoiding frequent heavy rain. 
Ethan opened the \textit{Seattle Weather Exploration} example on the official Vega-Lite website to find the ideal time through visual analysis.\footnote{\url{https://vega.github.io/vega-lite/examples/interactive_seattle_weather.html}}
The page contained an upper scatterplot showing date and maximum daily temperature, with precipitation encoded by point size, and a lower bar chart summarizing the number of records for each weather condition.
Without exporting the data or reconstructing the visualization, Ethan activated the WidgetVA browser extension on the existing page (R3) (\Cref{fig:official_page_case} A1--A2).
The two linked views were imported into a two-widget workspace.
Ethan then sent the question: ``We want Seattle's moody, misty atmosphere for an outdoor shoot, but frequent heavy rain would disrupt filming. Can you help identify a time that satisfies this need?'' (\Cref{fig:official_page_case} A2).}

\paragraph{\revise{Cross-Widget Exploration}}
\revise{The agent first translated the requested atmosphere into the relevant weather categories, fog and drizzle, and selected them in the bar chart (R1). The cross-view linkage filtered the scatterplot to the corresponding records, revealing several candidate periods from late summer to fall (\Cref{fig:official_page_case} B1). Through subsequent temporal brushing and comparison of the weather distributions, the agent progressively narrowed to September (\Cref{fig:official_page_case} B2).}
%: earlier periods contained substantially more sunny days, whereas later periods provided more fog and drizzle but also increased rain.

\paragraph{\revise{September Trade-Off}}
\revise{The agent divided September into two possible time periods and brushed Sept.~1--15 and Sept.~16--30 separately (\Cref{fig:official_page_case} C1--C2).
The linked bar chart showed that the first half had fewer rainy and foggy days.
The second half contained more fog and drizzle records, producing a stronger misty appearance.
However, rain also increased substantially and became almost as frequent as sun.
The comparison therefore isolated the remaining decision: Sept.~1--15 offered more reliable filming conditions, whereas Sept.~16--30 better matched the desired atmosphere but might impose greater disruption (\Cref{fig:official_page_case} C).}

\paragraph{\revise{Rain-Severity Verification}}
\revise{Weather composition alone did not reveal whether the additional rain in late September was likely to cause substantial disruption. The Interaction Trace preserved both September states, allowing the agent to revisit each interval. For each interval, the agent selected \texttt{rain} in the linked bar chart and inspected the remaining rainy-day records in the scatterplot (\Cref{fig:official_page_case} D1--D2).}
\revise{
The Sept.~16--30 state contained more rainy days, several larger precipitation marks, and lower rainy-day temperatures than the Sept.~1--15 state (\Cref{fig:official_page_case} D1--D2).
This evidence suggested that the stronger misty atmosphere in late September came with a higher risk of filming disruption.
Based on this additional evidence, Ethan selected Sept.~1--15 as the primary shooting period because it preserved some fog while providing fewer rainy days and better filming conditions. She left Sept.~16--30 as an alternative for scenes requiring a stronger misty atmosphere (\Cref{fig:official_page_case} D).}

\subsection{Failure Modes Analysis}
\label{sec:failure_modes}
% TODO: Define a failure taxonomy covering perception errors, action grounding errors, and planning drift.
% TODO: Provide case studies with step-by-step traces.
% TODO: Quantify error patterns when the paper reports them.
% TODO: Translate findings into implications for model and system design.

Beyond the aggregate scores, both the benchmark evaluation and the usage scenarios reveal several recurring failure patterns.
We report important ones here in the hope that they can help guide future research on agentic visual analytics.

\textbf{\revise{Evidence-to-answer grounding errors.}}
Although recent VLMs have made notable progress in visual analysis, they still sometimes \revise{derive wrong conclusions from the available evidence.}
\revise{Across the $S_3$ condition, this pattern---defined as Answer $<0.5$ with both Trace and State scores of at least $0.6$---occurred in 3.8\% of GPT episodes, 13.5\% of Gemini episodes, and 16.8\% of Qwen episodes.} 
\revise{For example, in a constrained-ranking task, GPT correctly filtered the requested categories, sorted the bars, highlighted the leading candidates, and compared the visible groups (Trace = State = 1.0), but answered \textit{West} when the correct answer was \textit{East}. In a cohort task that also exhibited planning drift, Qwen correctly brushed the requested region but returned only \textit{Male}, omitting the required comparison between \textit{Male} (36 records) and \textit{Female} (8 records). These examples show that successful interaction does not by itself guarantee that the model selects the correct visible statistic or answers the query correctly.}

\textbf{Tool-grounding errors.}
Tool-grounding errors arise when the model invokes tools, but not the ones that best match the user's analytical intent.
\revise{In the $S_3$ condition, 6.5\% of GPT episodes, 9.5\% of Gemini episodes, and 17.0\% of Qwen episodes had Answer, Trace, and State scores all below $0.5$. For example, in several tasks whose reference trace required brushing a specified price-revenue region, models applied only a categorical filter and then answered from the unbrushed view. The issue is therefore not tool availability, but grounding the semantic request that makes the relevant evidence observable.}

\textbf{Planning drift and non-convergent exploration.}
Planning drift appears when the model remains active in the tool loop but fails to converge on the interaction that actually resolves the task.
\revise{In one scatter-cohort task under $S_3$, Qwen exhausted the 16-turn interaction budget. It brushed the requested region and then issued 13 repeated \texttt{perception.summarizeVisible} calls, along with view-configuration checks. It ultimately returned only \texttt{Male}, omitting the required comparison between \texttt{Male}: 36 records and \texttt{Female}: 8 records. The trace suggests that stronger evidence-sufficiency and stopping criteria are needed.}

\section{Discussion and Limitations}
\label{sec:discussion_limitations}

\subsection{Discussion}
\label{sec:discussion}

% \revise{In addition to the user scenarios}, we conducted follow-up interviews with E1--E3 to collect their reflections on the system and widget abstraction. 
After the evaluation, we conducted follow-up interviews with E1--E3 to collect their reflections on \revise{\name}, \revise{which we report as formative reflections rather than as independent validation.}

\textbf{Observability and Provenance.}
Overall, the experts found our system easy to work with, not merely because it automates parts of the analysis, but because it makes the analytical process visible and inspectable.
E2 especially valued the Interaction Trace view, which helped him \textit{``understand the exploration logic and how the current state was reached''}.
E3 also added, \textit{``The transparency of the provenance makes the agent easier to follow.''}
These reflections suggest that transparency is not merely a presentation feature, but an important condition for making mixed-initiative collaboration feel controllable and comfortable.
\revise{In our framework, agent-issued actions are reflected in the shared widget and workspace states and recorded in the analysis history.} Users can therefore connect an intermediate view to the interaction that produced it, revisit previous states, and continue the analysis from an earlier point.
Therefore, the analytical process becomes explicit and replayable, making the co-analysis between human and agent more controllable and trustworthy.

\textbf{Beyond Model Capability.}
Our evaluation results suggest that effective agentic VA depends not only on model capability, \revise{but also on how the VA environment exposes interaction state and capabilities to the agent.}
\revise{Across the three evaluated model families, overall Answer performance generally follows the same ordering over the four execution settings. With executable capabilities fixed, Widget Abstraction further improves interaction grounding, while Workflow provides its strongest support when the analytical path is underspecified. The larger gains in Trace and State than in Answer for some models further indicate that interface-side support can improve evidence acquisition before those gains are fully translated into final conclusions.} 
\revise{Qualitatively, the usage scenarios and expert reflections further show that effective agentic VA requires interfaces that support observability, intervention, and continuation during analysis.}
These results imply that, alongside continued advances in model capability, more effort should also be devoted to designing systems that can better utilize those models for VA-assisted analysis. 
This includes not only stronger agentic system design, but also interface and interaction designs that better involve humans in practice.

\textbf{Diagnostic Benchmarking.}
As agentic systems continue to evolve rapidly, evaluation should do more than report end-task success.
Final accuracy alone provides limited insight into why an agent fails or which component has improved under a new model or scaffold.
In contrast, \revise{\bench} is designed to provide more diagnostic feedback by evaluating behavior along multiple dimensions, including Answer, \revise{Reference Trace Similarity}, and State, together with intermediate view states and reference traces.
This diagnostic capability makes it possible to localize bottlenecks more precisely, which helps researchers move beyond coarse comparison and more directly target the limitations of autonomous VA agents and their supporting systems.

\subsection{Limitations and Future Work}
\label{sec:limitations}
% TODO: State scope limitations for interfaces, datasets, and task coverage.
% TODO: State threats to validity such as simulator assumptions and prompt sensitivity.

We also identify several limitations and outline promising extensions to the framework and benchmark.

\textbf{Non-Unique Interaction Traces.}
In many cases, annotators recorded the interaction trace that an efficient human analyst would naturally take, such as invoking \texttt{\revise{bar.sortBar}} for a ranking task.
\revise{However, there can be multiple valid traces for the same analysis intent.
To reduce the dependence on a single reference trace, our current Reference Trace Similarity already supports annotated alternative steps and partial parameter matches. In addition, the separate State evaluation checks whether the required semantic conditions are satisfied in the resulting workspace, allowing successful trajectories that differ from
the primary reference trace to receive credit.}
\revise{Nevertheless, these mechanisms can only account for alternatives anticipated during benchmark annotation} and cannot exhaustively represent all valid analytical strategies or assess their relative quality.
Future benchmark designs should \revise{expand the coverage of alternative interaction paths and} provide multiple valid traces and graded preferences over interaction paths in terms of efficiency, interpretability, or analytical groundedness. 
This may also serve as a useful foundation for future RL-based optimization.

\textbf{\revise{Agent-Side Learning from Interaction Traces.}}
\revise{The evaluation still shows remaining agent failures in evidence-to-answer grounding, tool selection, and planning, even under $S_3$.
The present work diagnoses these gaps and studies interface-side support, but it does not specialize or train agents for visual analysis.
One obstacle is the scarcity of high-quality traces aligned with analytical intent, rather than with pointer events or screenshots.
Because human and agent operations already share the same widget state--action--perception contract, ordinary analysis sessions can be recorded as structured traces without a separate logging protocol.
Such traces could later be post-processed and annotated, then reused to construct memory and reusable skills or to train agents for VA tasks.
For remaining agent-side failures, these traces may be more directly useful than evaluation-oriented reference paths alone.}

\textbf{\revise{Data and Visualization Scope.}}
\revise{Our current framework and benchmark focus on structured tabular data and a controlled set of six registered widget families: bar charts, line charts, scatterplots, heatmaps, parallel coordinates, and Sankey diagrams. In particular, the current benchmark does not cover unstructured or multimodal data, whose use would require additional mechanisms for extracting task-relevant semantics and representing them through agent-operable states, actions, and perceptions. Future work should therefore extend both the widget library and benchmark to broader data representations and visualization forms~\cite{li2026NCP,zhou2024cluster,zhou2025hierarchical,li2025ruleexplorer}, and investigate whether the current abstraction remains sufficient as their interaction and state semantics become more heterogeneous.}

%Although the widget contract separates agent-facing analytical semantics from native visualization implementations, and the adapter architecture is designed to support additional visualization types, our empirical results do not establish coverage of arbitrary visualizations or VA systems.
\section{Conclusion}
\label{sec:conclusion}

We introduced \revise{\name}, a widget-centric framework for agentic VA.
By standardizing interactive visual components as structured widgets with unified action and perception interfaces, our framework makes agent interaction more observable and reusable.
Built on this abstraction, \revise{we construct \bench, which} enables fine-grained evaluation of interaction-centric VA behavior through answers, traces, and final states.
Experiments with state-of-the-art VLMs show that \revise{our framework provides an effective scaffold for agentic VA}, but also reveal substantial remaining challenges in grounding, state management, and multi-step analytical control. 
We hope this work provides a useful foundation for developing more capable and more collaborative agentic VA systems.

% \noindent\textbf{SUPPLEMENTARY}
%\section*{Supplemental Materials}
%The supplementary document includes (1) the prompting templates, mode-specific instructions, and structured output schemas used by the agent, (2) the chart-type skill prompts, (3) the specific action and perception primitives, (4) the benchmark dataset and evaluation protocol, and (5) a video of case studies and demo scenarios discussed in the paper.
%The codebase and implementation details will also be included in the https://anonymous.4open.science/r/VisAgentBenchmark-C235.

\bibliographystyle{IEEEtran}
\bibliography{main}

\end{document}